\documentclass[review]{elsarticle}

\usepackage{lineno,hyperref,amsmath,amssymb}
\modulolinenumbers[5]
\begin{document}

\begin{frontmatter}

\title{When diffusion faces drift: consequences of exclusion processes
   for bi--directional pedestrian flows}
\tnotetext[mytitlenote]{Fully documented templates are available in the elsarticle package on \href{http://www.ctan.org/tex-archive/macros/latex/contrib/elsarticle}{CTAN}.}

\author[mymainaddress]{Emilio N.M.\ Cirillo}
\ead{emilio.cirillo@uniroma1.it}

\author[mysecondaryaddress]{Matteo Colangeli}
\ead{matteo.colangeli1@univaq.it}

\author[mythirdaddress]{Adrian Muntean}
\ead{adrian.muntean@kau.se}

\author[myaddress,mythirdaddress]{T.K.\ Thoa Thieu}
\ead{thoa.thieu@kau.se}



\address[mymainaddress]{Dipartimento di Scienze di Base e Applicate per 
	l'Ingegneria, Sapienza Universit\`a di Roma, 
	via A.\ Scarpa 16, I--00161, Roma, Italy.}
\address[mysecondaryaddress]{Dipartimento di Ingegneria e Scienze dell'Informazione e
	Matematica, Universit\`a degli Studi dell'Aquila,
	via Vetoio, I--67100 L'Aquila, Italy.}
\address[mythirdaddress]{Department of Mathematics and Computer Science,
	Karlstad University, Sweden.}
\address[myaddress]{Department of Mathematics, Gran Sasso Science Institute, 
	Viale Francesco Crispi 7, 67100, L'Aquila, Italy.}

\begin{abstract}
Stochastic particle--based models are useful tools for describing
the collective movement of large crowds of pedestrians in crowded  confined environments. Using descriptions based on the simple exclusion process, two populations of particles, mimicking pedestrians walking in a built environment, enter a room from two opposite sides. One population is {\em passive} -- being unaware of the local environment; particles belonging to this group perform a symmetric 
random walk. The other population has information on the local geometry in the sense that as soon as particles enter a visibility zone, a drift activates them. Their self-propulsion   leads them towards the exit. This second type of species is referred here as {\em active}. The assumed crowdedness corresponds 
to a near--jammed scenario. The main question we ask in this paper is: {\em Can we induce modifications of the dynamics of the active particles to improve the outgoing current of the passive particles?}
To address this question, we compute 
occupation number profiles and 
currents for both populations in selected parameter ranges. Besides 
observing the more classical faster--is--slower effect, new features 
appear as prominent  like the  non--monotonicity of currents, 
self--induced  phase separation within the active population, 
as well as acceleration 
of passive particles for  large--drift regimes of active particles.  
\end{abstract}


\end{frontmatter}


\section{Introduction}
\label{s:int}
\par\noindent
Scenarios of pedestrian flows in agglomerated urban environments, 
cellular membranes, glasses, supercooled liquids share an 
important feature -- the dynamics takes place in a crowded environment 
with obstacles that are often active (i.e., not necessarily fixed in 
space and time). The management of the dynamics in these kinds of systems 
is far from being understood mainly due to the fact that the interplay 
between transport and particle--particle as well as particle--obstacle 
interactions is very complex; see, e.g., 
\cite{Ghosh2015,Simpson2009,Cirillo2019, Wang_PRL} 
and references cited therein.  

Exploring by means of computer simulations is an efficient 
tool for investigating the qualitative behavior of such  kinds of systems 
and this is also the route we take here. 

Depending on the level of observation, the modeling descriptions refer 
to micro--, meso--, macro--levels, or to suitable (multiscale) 
combinations thereof (see, e.g., \cite{WE,Muntean2014}). 
In this framework, we consider a microscopic approach based on a 
modification of the  classical simple 
exclusion\footnote{A simple exclusion process refers to the stochastic 
motion of interacting particles on a lattice where the interaction is 
given by the exclusion (excluded ``volume" constraints) property, i.e., 
two particles may not occupy the same site simultaneously. 
We refer the reader to \cite{Errico} for rigorous considerations on the 
simple exclusion model and its hydrodynamic limits 
and to \cite{Galanti} for a basic modeling perspective.}  
process formulated for two different populations of interacting 
pedestrians (particles).  
Essentially, two populations of particles, mimicking pedestrians walking 
in a built environment, enter a room from two opposite sides with the 
intention to cross the room and then exit from the door on the opposite 
side of the room.  

Because of its own unawareness or lack of prior knowledge of the local 
environment, one population is {\em passive}, and hence,  particles 
belonging to this population perform a random walk. 
On the other hand, the second  population has information on the local 
geometry in the sense that as soon as particles enter a visibility zone, 
a drift activates them by sending them towards the exit door. 
This  type of particles is referred here as {\em active}. 
The assumed crowdedness corresponds to a near--jammed scenario. 
To fix ideas, the number of occupied
sites in the room is chosen to be of the order of the $60\%$ from the
total number of available sites\footnote{Museums in highly touristic cities
are examples of crowded areas; 
compare, e.g., with the situation of Galleria Borghese in Rome as 
described in \cite{Corbetta}.}.

In an evacuation due to an emergency situation (like fire and smoke 
propagating in the building), the {\em a priori} knowledge of the 
environment is certainly an advantage (cf., e.g., \cite{Omar}). 
Hence, from this perspective, if a quick evacuation is needed, 
then the passive population has a disadvantage compared to the 
active population.  
We are wondering whether we can compensate at least partly this drawback, 
by managing intelligently the motion of the active population. 
In other words, 
the main question we ask in this paper is: 
\vskip 0.5 cm

\fbox{
\parbox{0.8\textwidth}{
Q.\ Can we induce modifications of the dynamics of 
active particles to improve the outgoing current of 
passive particles?}}

\vskip 0.5 cm
The ingredients we have at our disposal are alterations 
either in the drift parameter of the active particles, 
or in their visibility zone by fine--tuning a parameter for a nonlocal 
interaction that activates the drift--towards--exit. 
It is worth noting that the 
latter feature is different from the nonlocal shoving 
of particles proposed in \cite{Landman_shoving}. 

To address the above question, we compute occupation number profiles 
and currents for both populations in selected parameter ranges. 
Our numerical results exhibit the classical faster--is--slower 
effect (see, e.g., \cite{Garcimartin} for experimental evidence 
and \cite{Tomoeda} for numerical simulations exhibiting this effect 
using Helbing's social force model) and point out as well new  
prominent features like the non--monotonicity of currents, 
the self--induced phase separation within the active population, 
as well as an acceleration of passive particles induced by  
a large drift and a large visibility zone  of active particles.  

This research was initiated in \cite{Cirillo2016,Cirillo2016pre},  
motivated by our intention to estimate the mean residence time of 
particles undergoing an asymmetric simple exclusion within a room 
in perfect contact with two infinite reservoirs of particles. 
Recent developments reported in \cite{Cirillo2019} brought us to study 
drafting effects via the dynamics of mixed active--passive pedestrian 
populations in confined domains with obstacles and exit doors, 
which mimick a built complex environment.  

\section{Model description}
\label{s:mod}
\par\noindent
We consider a square 
lattice $\Lambda:=\{1,\dots,L\}\times\{1,\dots,L\}\subset \mathbb{Z}^2$ of side 
length $L$, where $L$ is an odd positive integer number. $\Lambda$ will be referred in this context as {\em room}.  
An element $x=(x_1,x_2)$ of the room $\Lambda$ is called \emph{site} 
or \emph{cell}.
Two sites $x,y\in\Lambda$ are said \emph{nearest neighbor} if and only 
if $|x-y|=1$.
Let us call \emph{horizontal} the first coordinate 
axis and \emph{vertical} the second one. The words 
\emph{left}, 
\emph{right}, 
\emph{up}, 
\emph{down}, 
\emph{top}, 
\emph{bottom}, 
\emph{above}, 
\emph{below}, 
\emph{row}, 
and 
\emph{column} 
will be used accordingly. We call \emph{door} the sets make of 
$w_{\textup{L}}$ and 
$w_{\textup{R}}$ pairwise adjacent in the 
left--most and right--most columns of the room $\Lambda$, respectively, 
symmetric with respect to its median row. 
This mimics the presence of two distinct doors on the left and 
right boundary of the room. The odd positive integers  
$w_{\textup{L}},w_{\textup{R}}$ smaller than $L$ 
will be called \emph{width} of the doors.
Inside the  
room we define a rectangular driven zone, namely, 
the \emph{visibility region} $V$, made of the 
first $L_{\textrm{v}}$ left columns of $\Lambda$, with the positive integer 
$L_{\textrm{v}}\le L$ called \emph{depth} of the visibility region.
By writing $L_{\textrm{v}}=0$, we refer to the case in which no 
visibility region is considered. 

We  consider two different species of particles, i.e.,
\emph{active} and \emph{passive}, moving inside $\Lambda$
(we shall sometimes use in the notation the symbols A and P to refer to them).
The dynamics will be defined so that the sites 
of the external boundary of the room, that is to say the sites  
$x\in \mathbb{Z}^2\setminus\Lambda$ such that there exists 
$y\in\Lambda$ nearest neighbor of $x$,
cannot be accessed by the particles.
The state of the system is a 
\emph{configuration} $\eta\in\Omega=\{-1,0,1\}^\Lambda$ 
and 
we say that the site $x$ is 
\emph{empty} if $\eta_x=0$,
\emph{occupied by an active particle} if $\eta_x=1$,
and
\emph{occupied by a passive particle} if $\eta_x=-1$.
The number of active (respectively, passive) 
particles in the configuration $\eta$ 
is given by 
$n_{\textrm{A}}(\eta)=\sum_{x\in\Lambda}\delta_{1,\eta_x}$
(respectively, $n_{\textrm{P}}(\eta)=\sum_{x\in\Lambda}\delta_{-1,\eta_x}$), 
where $\delta_{\cdot,\cdot}$ is Kronecker's symbol.
Their sum is the total number of particles in the configuration $\eta$.

\begin{figure}[h!]
	\centering
	\includegraphics[width=0.4\textwidth]{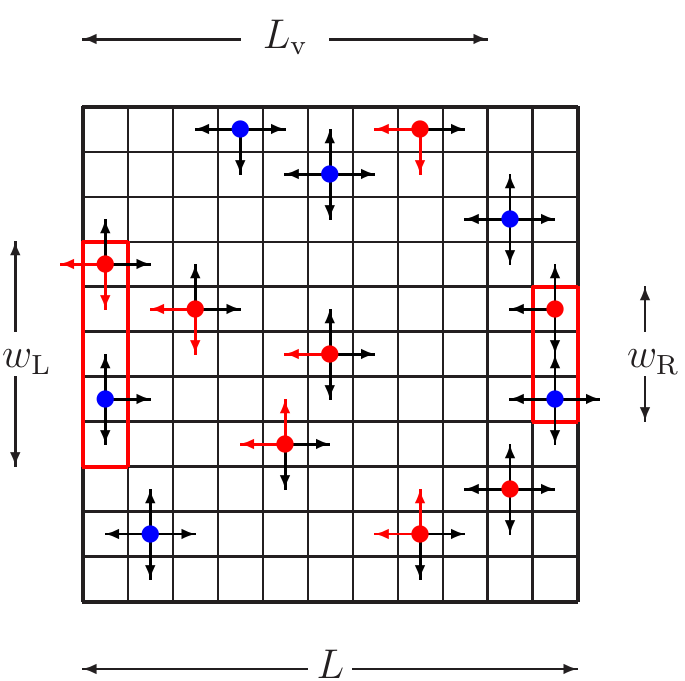}
	\caption{Schematic representation of our lattice model. Blue and red 
		disks denote passive and active particles, respectively. 
		The rectangles 
of sites delimited by the red contour denote the exit doors.
		Black and red arrows (color online) denote
		transitions performed with rates $1$ and $1+\varepsilon_1$ 
		or $1+\varepsilon_2$, respectively.}
	\label{fig:fig0}
\end{figure}

The interaction of particles inside the room is modeled via a 
simple exclusion random walk for  two particle species  undergoing 
two different microscopic dynamics. The passive particles enter through the left door and exit through the right door. They perform a 
symmetric simple exclusion dynamics on the whole lattice. Simultaneously, the active particles enter through the right door and exit through the left door. They perform a symmetric simple exclusion walk 
outside the visibility region, whereas inside such a region they 
experience also a drift pushing them towards the left door. 
In other words, the whole room is {\em obscure}\footnote{We refer the reader to \cite{Cirillo2016}, where the authors discuss the gregarious behavior of crowds moving in the dark.} for the passive 
particles, while, for the active ones, only the region outside the 
visibility region\footnote{The concept of visibility region was introduced by the authors in \cite{Cirillo2019}. } is obscure. 
The model also includes
two external particle waiting lists, each of which is designed to collect
particles of a given species when these move out from the lattice $\Lambda$
through their exit door and to reinsert them back on the lattice
through their entrance door. 

More precisely, 
picked the positive integers $N_\textup{A}$ and $N_\textup{B}$ and
set $N=N_\textup{A}+N_\textup{P}$, 
the dynamics is the continuous time Markov chain $\eta(t)$ on 
$\Omega$ 
with initial configuration $\eta(0)$ such that 
$n_\textup{A}(\eta(0))=N_\textup{A}$
and 
$n_\textup{P}(\eta(0))=N_\textup{P}$
and 
rates $c(\eta,\eta')$ defined as follows:
Let $\varepsilon_1,\varepsilon_2\ge0$ be the \emph{horizontal} and 
\emph{vertical drift};
for any pair $x=(x_1,x_2),y=(y_1,y_2)$ of nearest neighbor sites in $\Lambda$ 
we set $\epsilon(x,y)=0$, excepting the following cases:
\begin{itemize}
	\item[--]
        $\epsilon(x,y)=\varepsilon_1$
        if
	$x,y\in V$ and $y_1=x_1-1$, namely, $x$ and $y$ belong to the 
	visibility region and $x$ is to the right with respect to $y$;
	\item[--]
        $\epsilon(x,y)=\varepsilon_2$
        if
	$x,y\in V$ and $y_2=x_2+1\leq (L+1)/2$, namely, 
	$x$ belongs to the bottom part of the visibility region 
	and $x$ is below $y$;
	\item[--]
        $\epsilon(x,y)=\varepsilon_2$
        if
	$x,y\in V$ and $y_2=x_2-1\geq(L+1)/2$, namely, 
	$x$ belongs to the top part of the visibility region 
	and $x$ is above $y$.
\end{itemize}
Next, we let 
the rate $c(\eta,\eta')$ be equal 
\begin{itemize}
	\item[--]
	to
	$1$
	if $\eta'$ can be obtained by $\eta$ by 
	replacing with $0$ a $-1$ at the right door
	(passive particles leave the room);
	\item[--] 
	to
	$1+\epsilon(x,y)$
	if $\eta'$ can be obtained by $\eta$ by 
	replacing with $0$ a $1$ at the left door
	(active particles leave the room);
	\item[--]
	to $[N_{\textrm{A}}-n_{\textrm{A}}(\eta)]
	/m_\textup{R}$ 
if the number of empty sites in the right door is $m_\textup{R} >0$ 
and $\eta'$ can be obtained by $\eta$ by adding a $1$ at one of the empty sites of the right door;
	\item[--]
	to $[N_{\textrm{A}}-n_{\textrm{A}}(\eta)]
	/m_\textup{L}$ if the number of empty sites in the left door 
is $m_\textup{L} >0$ and $\eta'$ can be obtained by $\eta$ by adding a $-1$ at one of the empty sites of the left door;
	\item[--]
	to
	$1$
	if $\eta'$ can be obtained by $\eta$ by 
	exchanging a $-1$ with a $0$ between two neighboring sites of $\Lambda$ 
	(motion of passive particles inside $\Lambda$);
	\item[--]
	to
	$1+\epsilon(x,y)$
	if $\eta'$ can be obtained by $\eta$ by 
	exchanging a $+1$ at site $x$ with a $0$ at site $y$, 
	with $x$ and $y$ nearest neighbor sites of $\Lambda$ 
	(motion of active particles inside $\Lambda$);
	\item[--]
	to
	$0$ in all the other cases.
\end{itemize}

We stress that, 
at time $t$, the quantities 
$N_{\textrm{A}}-n_{\textrm{A}}(\eta(t))$
and 
$N_{\textrm{P}}-n_{\textrm{P}}(\eta(t))$
represent the number of active, and, respectively, passive 
particles 
that exited the room and entered their own waiting list at time
$t$, whereas $m_\textup{L}>0$ and $m_\textup{R}>0$
are the number of empty sites of the left and right doors at time $t$.

The system will reach a stationary state, since passive particles exiting 
the domain via the right door are introduced back in one site randomly 
chosen among possible empty sites of the left door, while active particles 
leaving the system through the left door are introduced back also at one 
random site chosen among possible empty sites of the right door. 
The total number of active and passive particles in the room $\Lambda$ 
is only approximatively 
constant during the evolution. It slightly fluctuates 
due to the fact that particles 
may enter waiting lists. On the other hand, the total number 
of particles $N$ 
in the system (considering both the room and the 
waiting lists) is conserved.

In the study of this dynamics, the main quantity of interest are the 
stationary \emph{outgoing fluxes} or \emph{currents
of active and passive particles}
which are the values approached in the infinite time limit 
by the ratio between the total number of active and 
passive particles, respectively,
that in the interval $(0,t)$
exited through the left and the right door 
and entered the waiting lists and the time $t$.
In order to discuss and to understand the behavior of currents with 
respect to the model parameters, we shall also look at the 
active and passive particles
\emph{occupation number profiles}, namely, we evaluate the stationary 
mean value of the occupation numbers 
$\delta_{1,\eta_x}$ 
and 
$\delta_{-1,\eta_x}$ 
of each site $x\in\Lambda$, where, we recall, $\delta_{\cdot,\cdot}$ is the 
Kronecker symbol.

\section{Numerical results}
\label{s:res}
\par\noindent
We simulate the model introduced in Section~\ref{s:mod}
using the following scheme:
at time $t$,  we 
extract an exponential random time $\tau$ with parameter
the total rate
$\sum_{\zeta\in\Omega}c(\eta(t),\zeta)$
and set the time equal to $t+\tau$.
We then select a configuration using the probability 
distribution 
$c(\eta(t),\xi)/\sum_{\zeta\in\Omega}c(\eta(t),\zeta)$
and then set $\eta(t+\tau)=\xi$. 

We compute the currents by applying directly the definition
given at the end of Section~\ref{s:mod} and the occupation number profiles
as follows: 
we  run the dynamics for a sufficiently 
long time (order of $9\times10^7$ MC steps)
so that the system reaches the stationary state and then we average
the occupation numbers 
$\delta_{1,\eta_x}$ 
and 
$\delta_{-1,\eta_x}$ 
for each site of the room
for the following $9\times10^7$ MC steps; we thus obtain 
a function of $x\in\Lambda$ 
taking values in the interval $[0,1]$.

The presence of two species fighting to get to opposing 
exit doors reasonably reduces the intensity of the current
that we would expect to measure if a single species were 
present in the room. We have tested this fact by running 
simulations for one single species with the 
some dynamics as the one defined above and with the 
same choice of parameters and we have found 
currents that are typically three or four time larger. 

In the following discussion the parameters are
$L=30$ and $N_{\textrm{A}}=N_{\textrm{P}}=280$
and
all of the simulations are done starting the system from the same 
initial configuration chosen once for all by distributing the particles 
at random with uniform probability. 
With such a choice of the parameters the number of occupied 
sites in the room is of the order of the $60\%$ of the 
total; indeed, our study, as explained in the introduction, 
aims at understanding the behavior of the model in a crowded regime.

\subsection{The corridor model}
\label{s:mod1}
\par\noindent
In this subsection, we consider the model defined above assuming 
that the doors are as wide as the room, namely, 
$L=w_{\textup{L}} = w_{\textup{R}}$.
In such case, it is rather natural to limit the discussion 
to the case $\varepsilon_2=0$, that is to say, to the case 
in which only the horizontal (longitudinal with respect to the 
position of the doors)
component of the drift is considered. 
To simplify the notation, we shall also 
denote the 
longitudinal drift
$\varepsilon_1$ 
simply by 
$\varepsilon$.

\begin{figure}[t]
	\centering
	\begin{tabular}{ll}
		\includegraphics[width=0.48\textwidth]{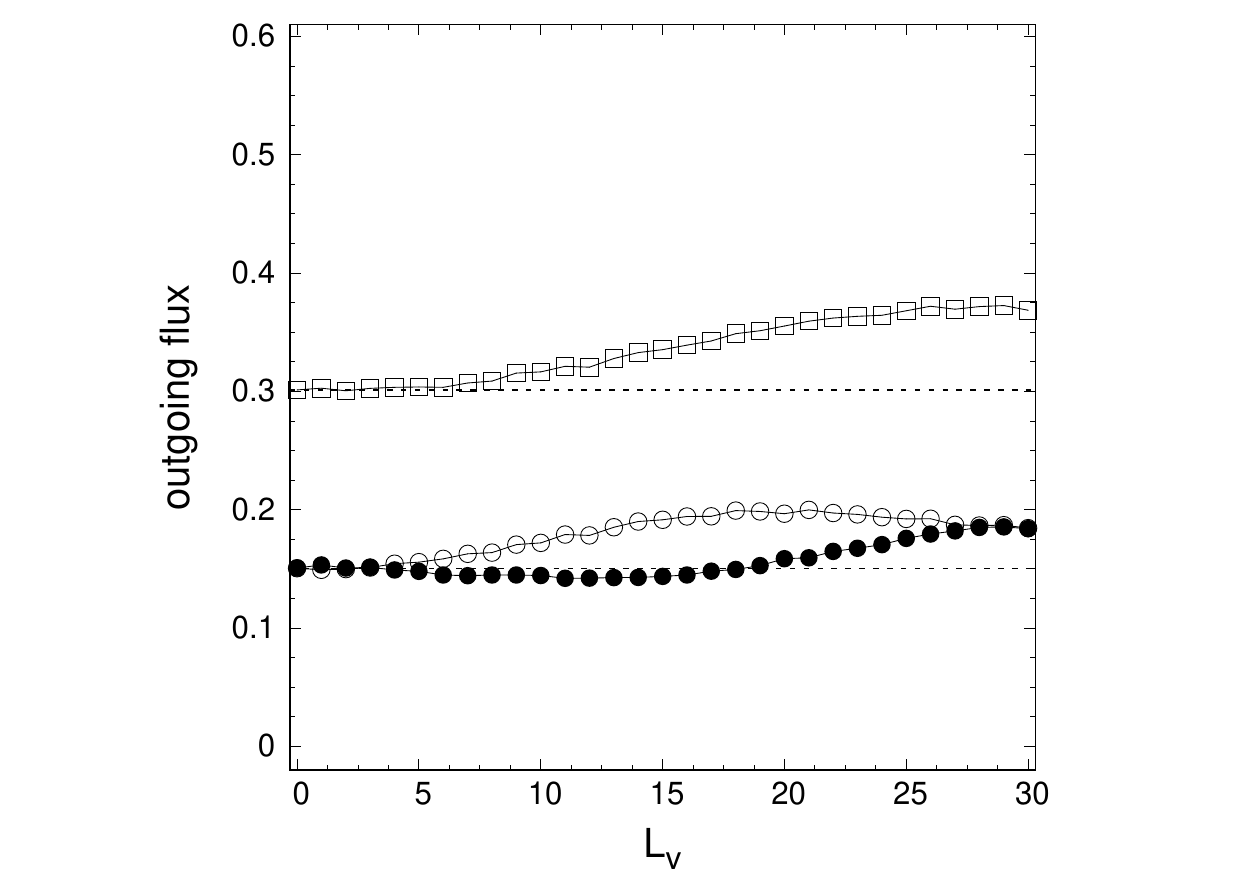}&
		\includegraphics[width=0.48\textwidth]{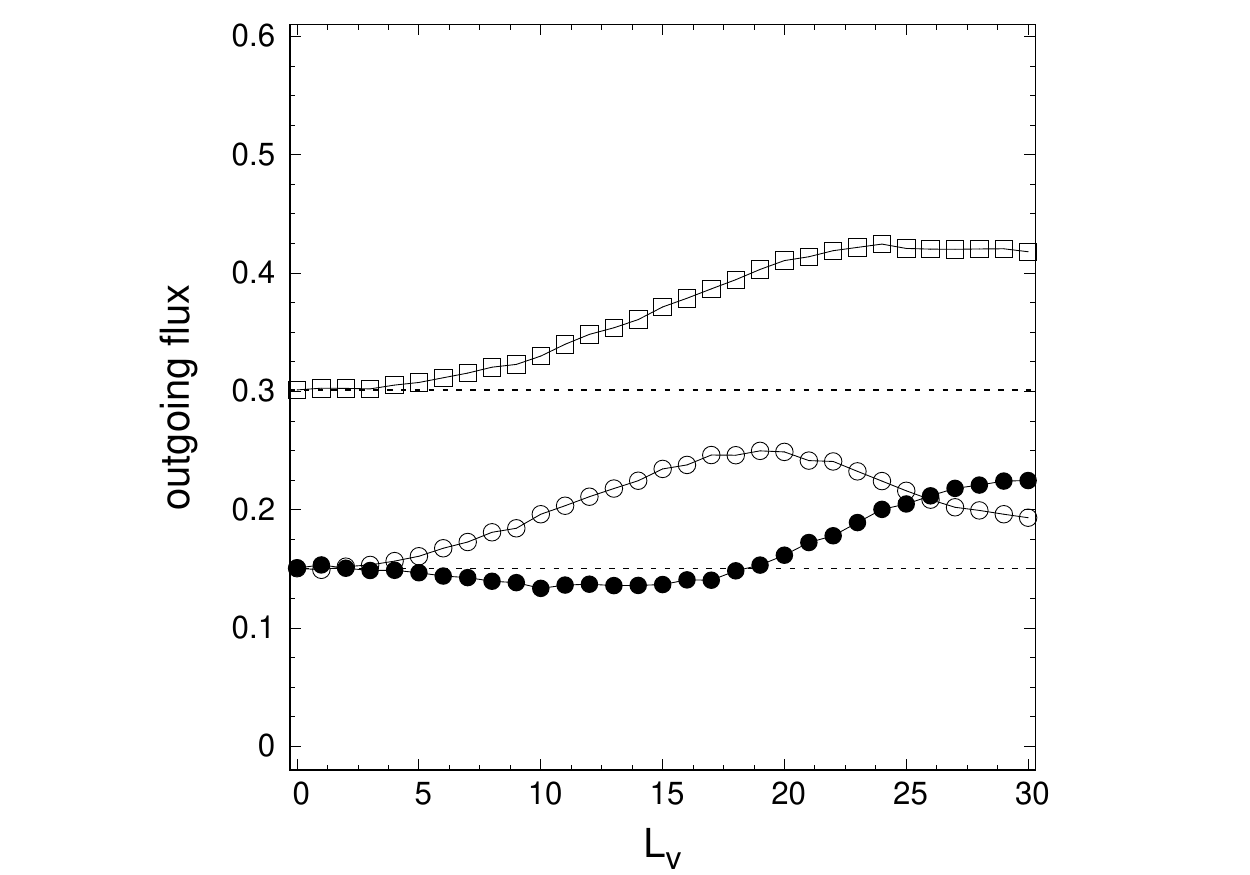}\\[0.1cm]
		\includegraphics[width=0.48\textwidth]{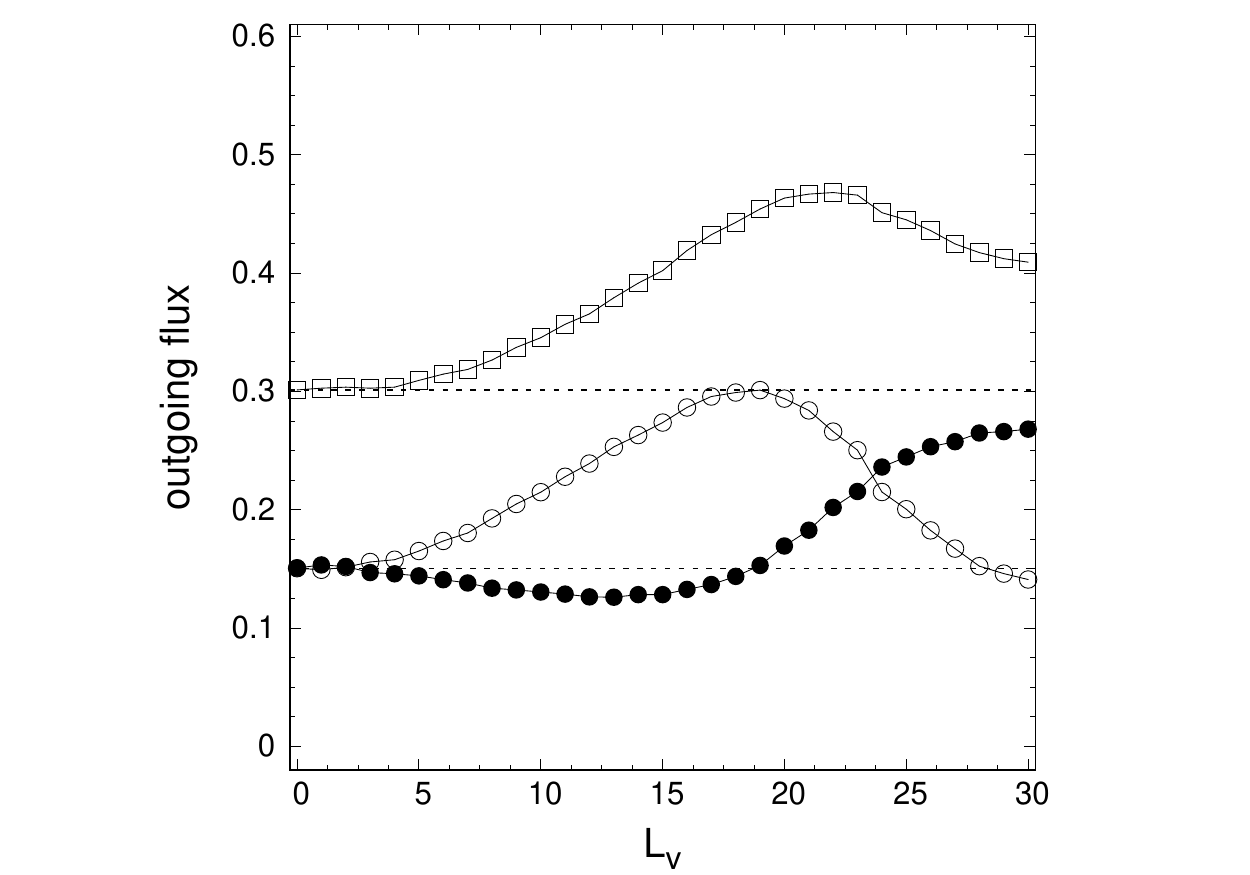}&
		\includegraphics[width=0.48\textwidth]{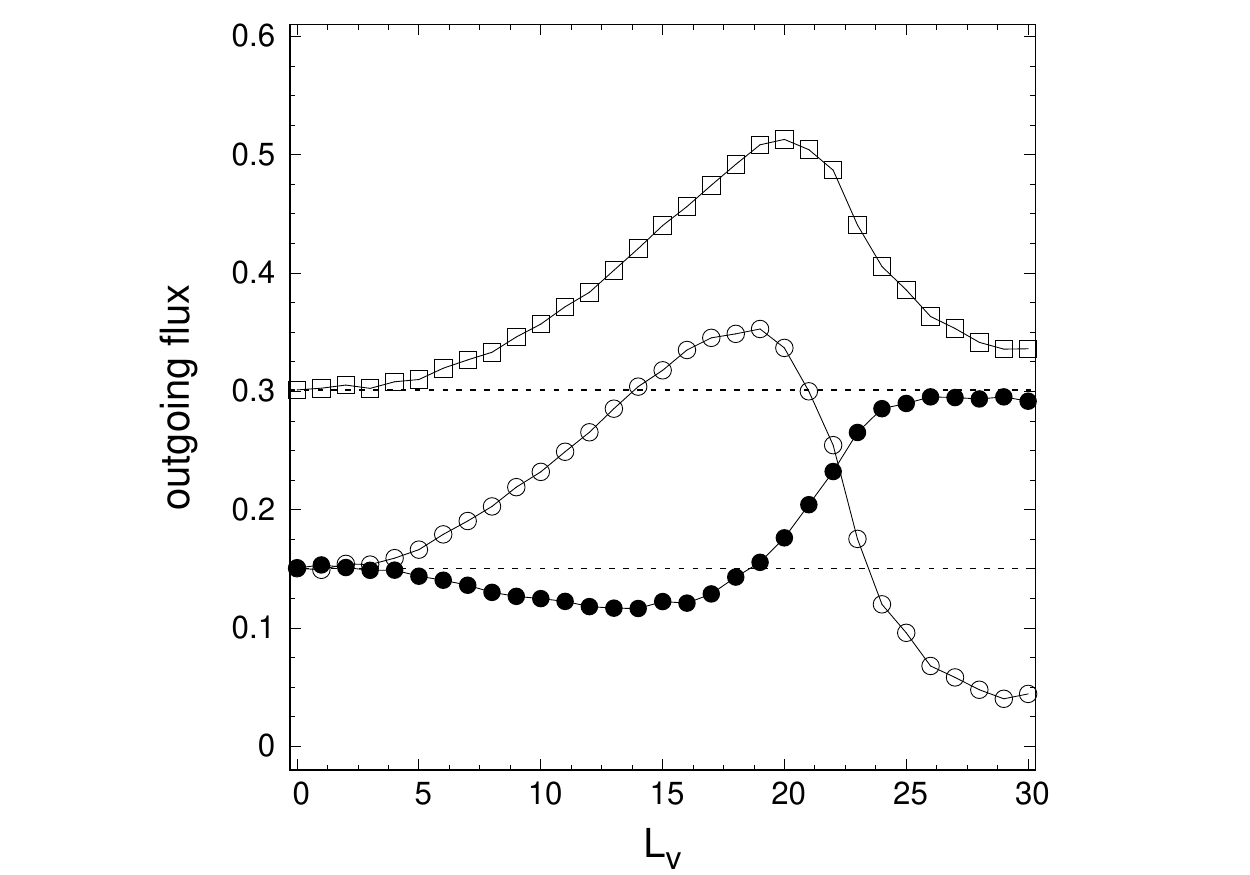}	
	\end{tabular}
	\caption{Stationary currents of active (empty circles) and passive particles (solid disks) and cumulative current (empty squares) as functions of $L_{\textrm{v}}$ 
		for $\varepsilon=0.05, 0.1,0.15,0.2$ (lexicographical order).  
The black dashed lines are eye guides showing the value 
measured in the zero drift case.
	}
	\label{fig:fig1}
\end{figure}

\begin{figure}[t]
\centering
\begin{tabular}{lll}
\includegraphics[width = 0.25\textwidth]{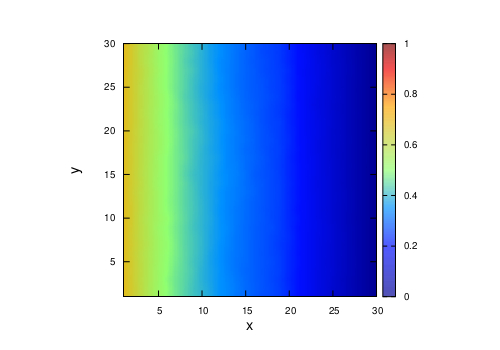}&
\includegraphics[width = 0.25\textwidth]{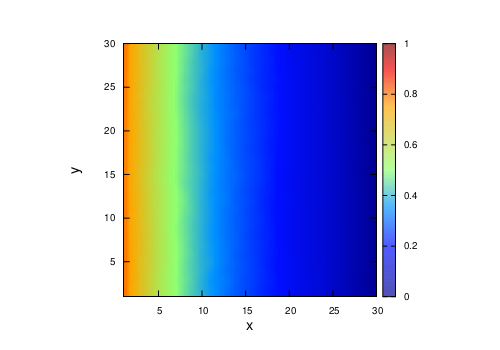}&
\includegraphics[width = 0.25\textwidth]{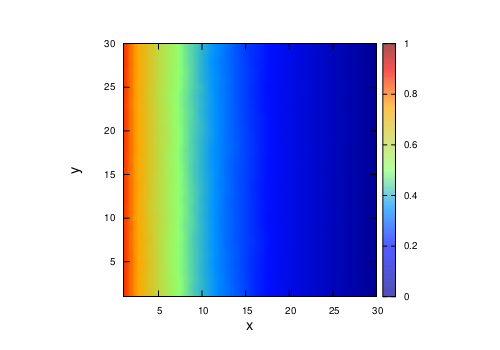}  
\\[0.1cm]
\includegraphics[width = 0.25\textwidth]{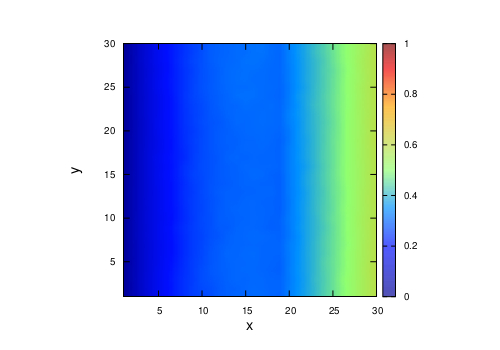}&
\includegraphics[width = 0.25\textwidth]{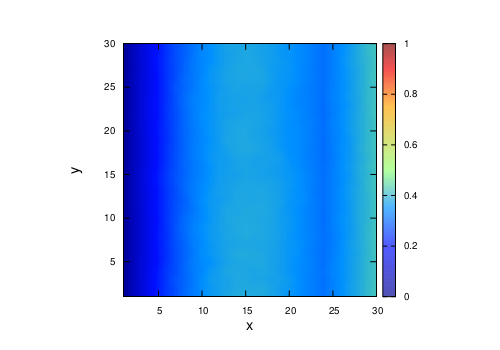}&
\includegraphics[width = 0.25\textwidth]{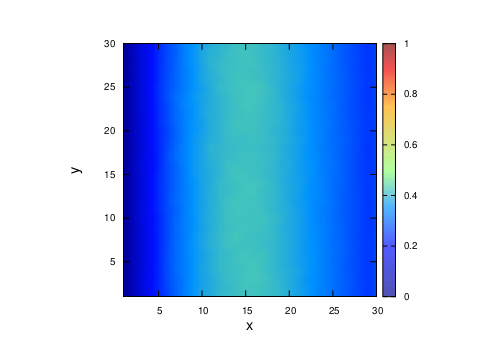}	
\end{tabular}
\caption{Occupation number profile of passive (top row) and active (bottom row) particles at stationarity for $\varepsilon=0.15$ 
and $L_{\textrm{v}}=20,25,30$ (from left to right).}
\label{fig:fig10}
\end{figure}

Among the different interesting results that we will discuss in 
this section, we want to 
single out a very peculiar behavior:
since only active particles experience the drift, 
changes of the parameters $L_{\textrm{v}}$ and $\varepsilon$
act directly only on the active species.
Nevertheless, due to the exclusion interaction between the 
two different species, also the passive particles behavior will 
be affected. In particular, we will see that a robust increase 
of the visibility length $L_{\textrm{v}}$ and 
the longitudinal drift $\varepsilon$ will induce
a significant increase of the passive particle current.

In Figure~\ref{fig:fig1}, we plot the passive and active particle 
currents for four different values of the drift, i.e.,
$\varepsilon=0.05, 0.1,0.15,0.2$, when the 
visibility length $L_{\textrm{v}}$ is varied from $0$ to $L$. 
We note that the active particles current (open circles) increases 
with $L_{\textrm{v}}$ up to some value where it attains a maximum. 
This effect is visible in all of the panels of Figure~\ref{fig:fig1}, 
but the position and size of the maximum changes. 
This effect is more prominent for the largest value of 
considered $\varepsilon$ (see bottom right panel in Figure~\ref{fig:fig1}). 
Moreover, excepting the smallest value of the 
drift $\varepsilon=0.05$, soon after the active particle 
current reaches its maximum, the passive and active particle 
currents intersect so that, for the largest value of the 
visibility length, the transport of passive particles becomes 
more efficient than that of active ones. 

This behavior is interesting and not trivial for two reasons: 
i) only active particles are driven and, hence, the parameters we play
with directly act only on their dynamics;
ii) when $\varepsilon$ and $L_\textup{v}$ are increased, the active 
particle transport throughout the corridor is expected to become more 
and more efficient.
To explain such an effect,
we look at the stationary occupation profiles.
In particular, we focus our attention to the case $\varepsilon=0.15$, 
namely, 
we closely analyze the 
bottom left panel in Figure~\ref{fig:fig1} and the corresponding 
occupation number profiles plotted in Figure~\ref{fig:fig10}. 
In this figure, we have considered the cases 
$L_\textup{v}=20,25,30$ since the switch in the active and passive 
particle currents is observed around 
$L_\textup{v}=25$.

Looking at Figure~\ref{fig:fig10}, we note that for 
$L_{\textrm{v}}=20$, passive and active particles mainly 
distributed close to their relative entrance doors, namely, 
passive particles on the left and active particle on the right. 
Nevertheless, a small depletion layer in the 
active particle distribution can be observed around 
$x=20$, which is precisely the place where 
those particles enter the visibility regions.
This is an expected behavior: active particle entering 
such a region from the right are accelerated towards the left and 
start to accumulate when they meet the passive particles standing 
at their left entrance. 
This behavior becomes more and more prominent when 
the visibility length is increased; see the panels 
corresponding to 
$L_{\textrm{v}}=25,30$, where the presence of the depletion 
region and the accumulation of the active particles in the middle 
of the room is more evident. 
As a consequence, when the visibility length is increased, 
the right entrance is no more occupied by active particles.
This explains the observed increase in the passive particle 
current.

\begin{figure}[t]
\centering
\begin{tabular}{ll}
\includegraphics[width=0.48\textwidth]{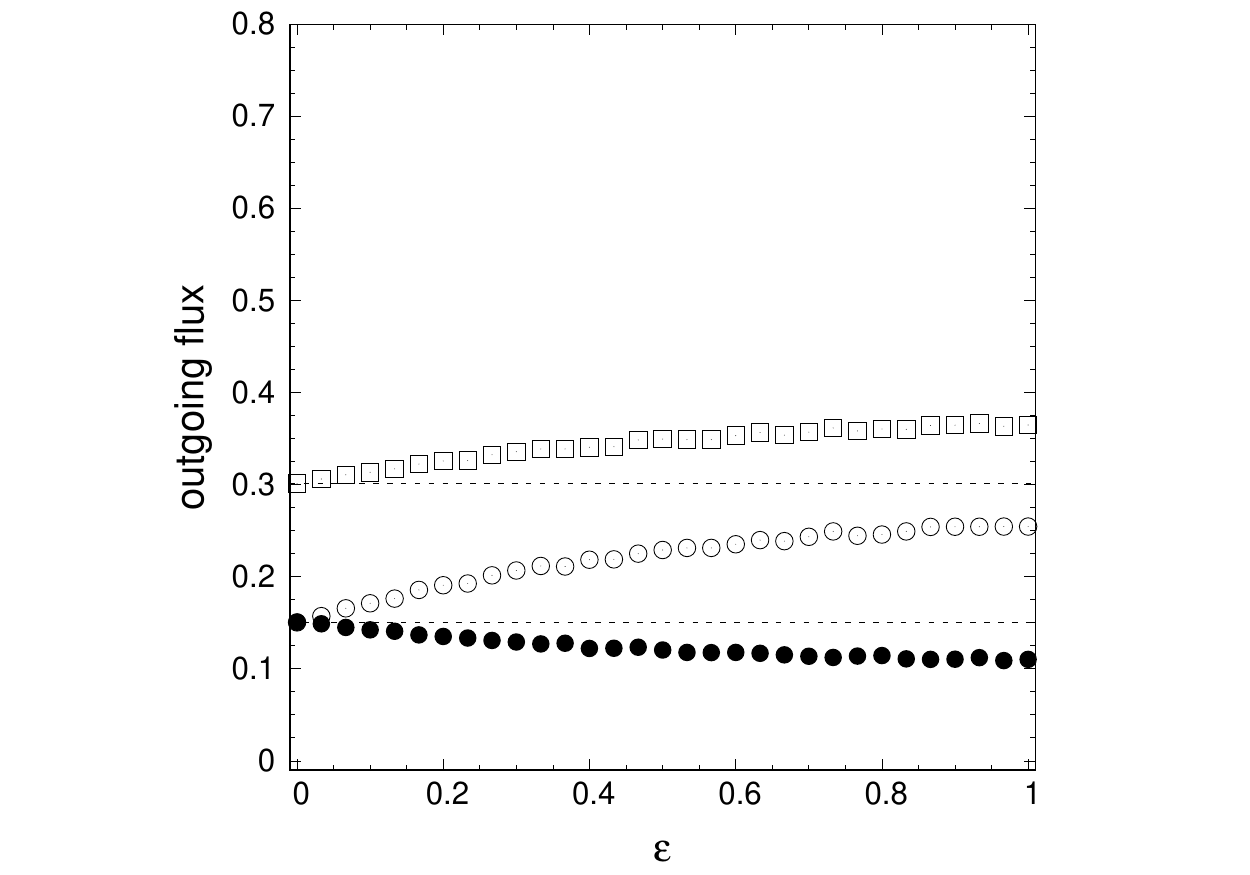}&
\includegraphics[width=0.48\textwidth]{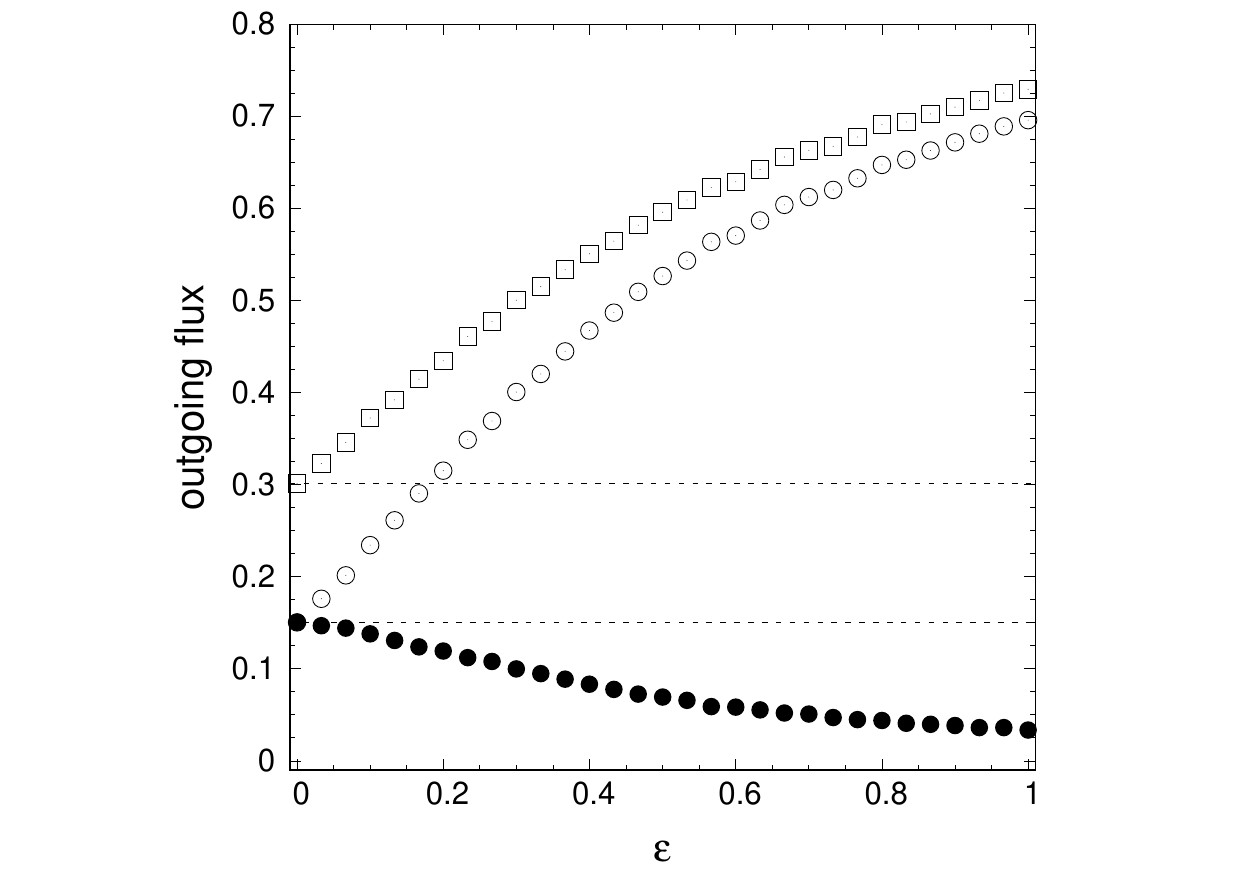}\\[0.1cm]
\includegraphics[width=0.48\textwidth]{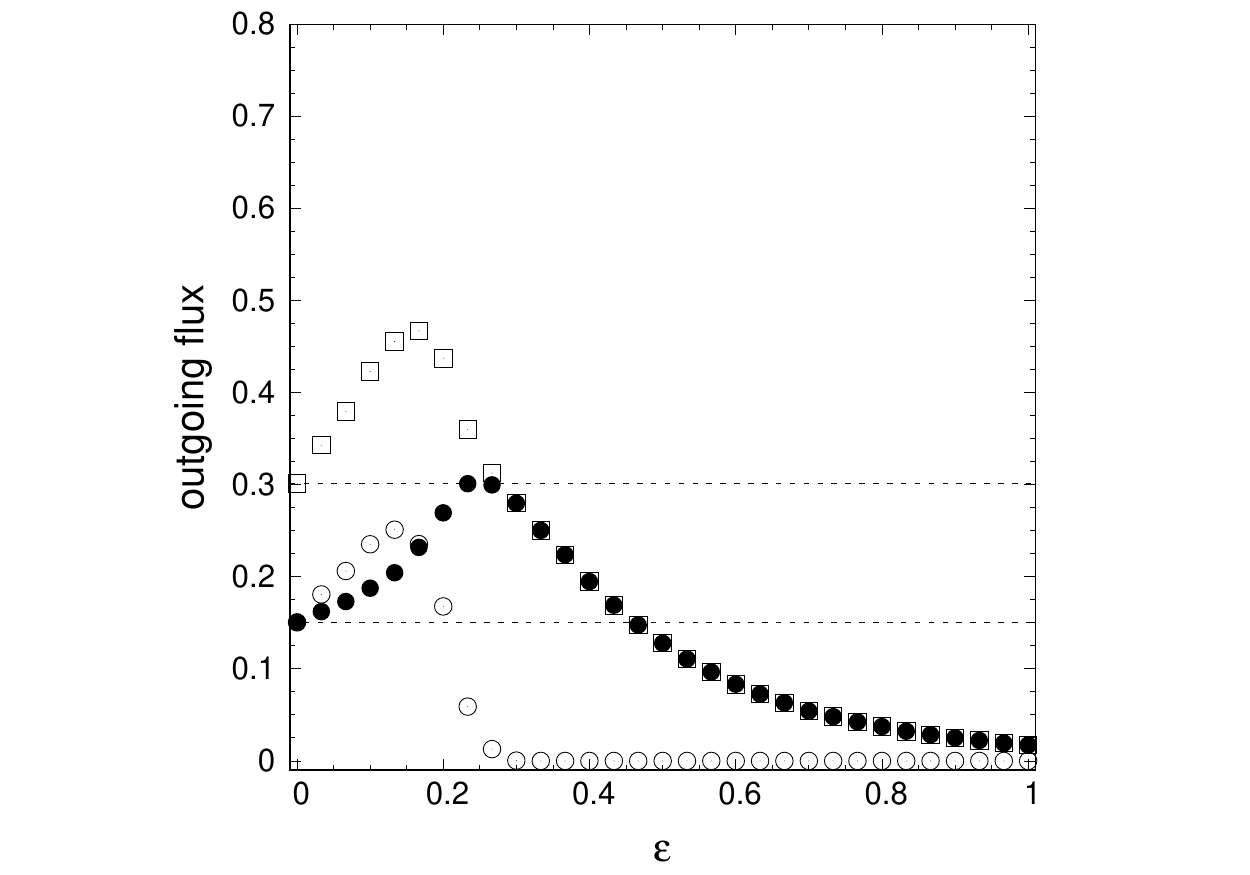}&
\includegraphics[width=0.48\textwidth]{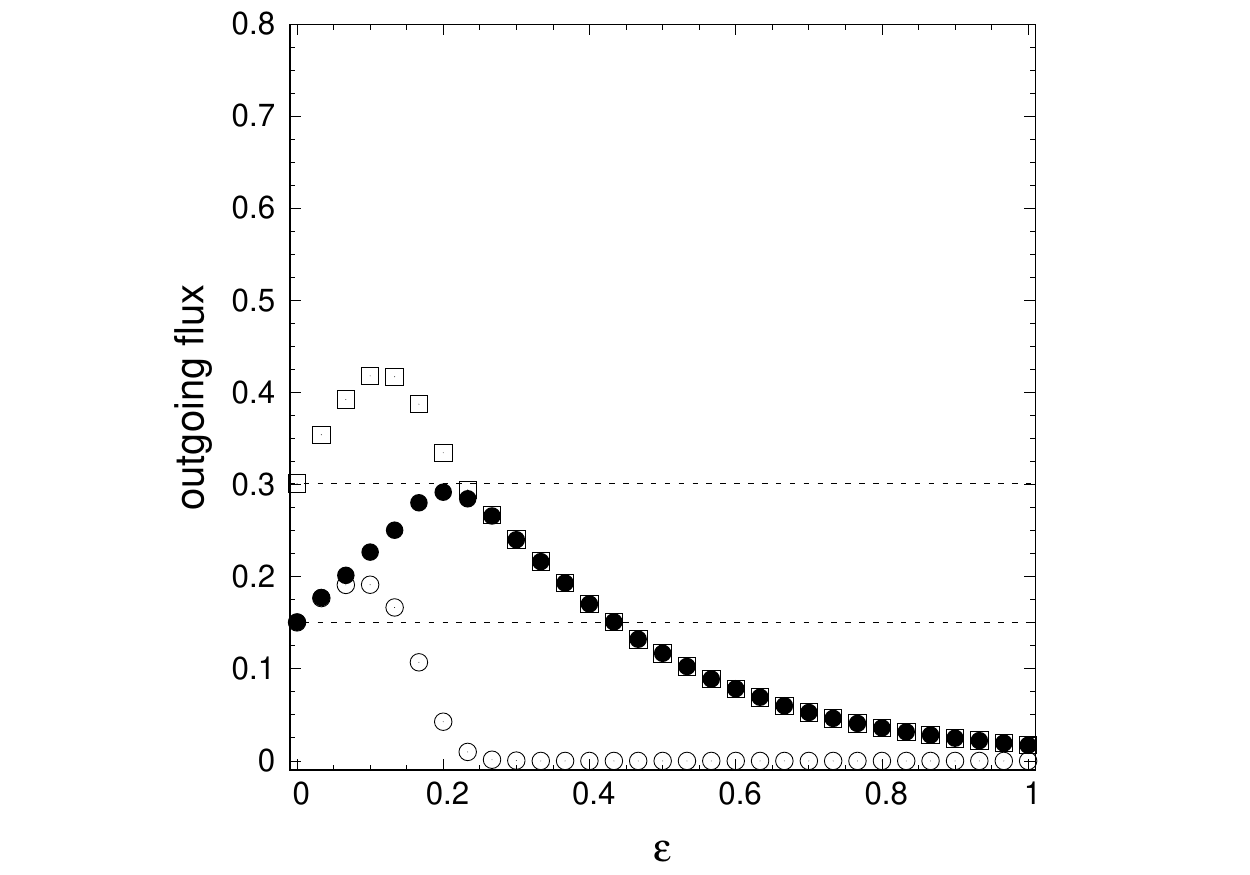}	
\end{tabular}
\caption{Stationary currents of active (empty circles) and passive particles (solid disks) and cumulative current (empty squares) as functions of $\varepsilon$ for $L_{\textrm{v}}=7,15,23,30$ (lexicographical order).  
The black dashed lines are as in Figure~\ref{fig:fig1}
}

\label{fig:fig3}
\end{figure}

Similar occupation profiles are found for the other values 
of the longitudinal drift considered in Figure~\ref{fig:fig1}.
Hence, 
the behavior of currents can be explained similarly. We do not 
report such pictures because they do not add anything new to 
the understanding 
of the behavior of our model.

\begin{figure}[t]
\centering
\begin{tabular}{lll}
\includegraphics[width = 0.25\textwidth]{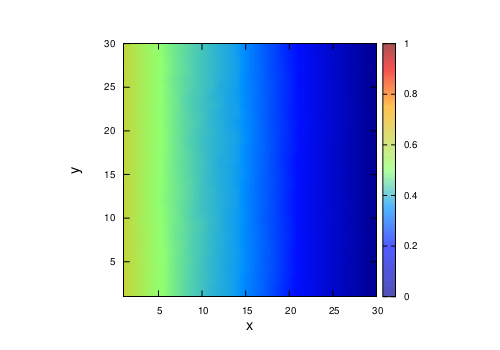}&
\includegraphics[width = 0.25\textwidth]{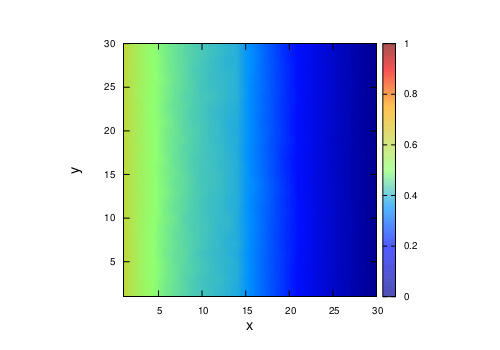}&
\includegraphics[width = 0.25\textwidth]{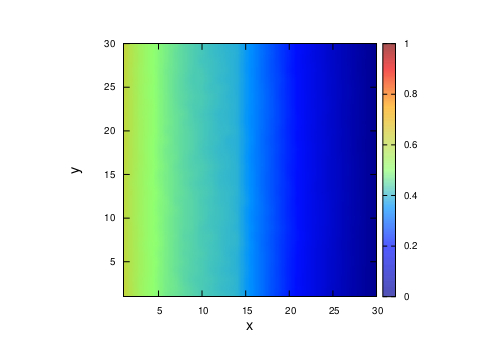}  
\\[0.1cm]
\includegraphics[width = 0.25\textwidth]{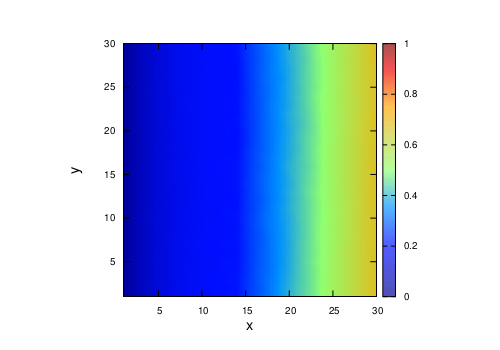}&
\includegraphics[width = 0.25\textwidth]{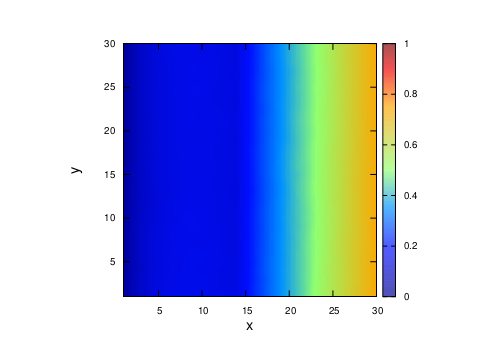}&
\includegraphics[width = 0.25\textwidth]{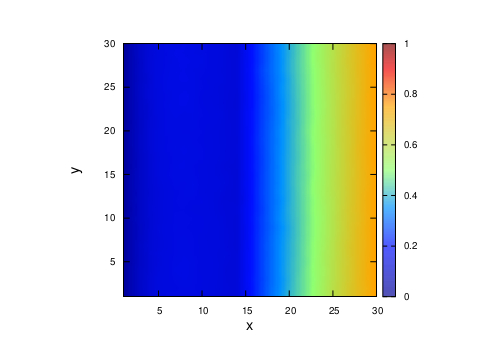} 	
\end{tabular}
\caption{Occupation number profile of passive (top row) and active (bottom row) particles at stationarity for  $L_{\textrm{v}}=15$ and $\varepsilon=0.2,0.4,0.48$ (from left to right).
}
\label{fig:fig12}
\end{figure}

In Figure \ref{fig:fig3}, we consider larger values of $\varepsilon$ 
for a fixed visibility length $L_\textup{v}=7,15,23,30$. 
If $L_{\textrm{v}}$ is smaller than $15$, the passive and, respectively,
active particle currents decrease and increase monotonically 
with respect to the drift $\varepsilon\in[0,1]$.
Moreover, to a high current of active particle it corresponds 
a small current of passive ones. 
This behavior is coherent with the occupation number profiles 
reported in Figure~\ref{fig:fig12}. We see an accumulation 
of active particles at the right door, which prevents high 
passive particle currents, as well as a rather spread distribution 
of passive particles in the left part of the room with minor accumulation 
at the entrance, which 
allows high active particle currents. 

\begin{figure}[t]
\centering
\begin{tabular}{llll}
\includegraphics[width = 0.25\textwidth]{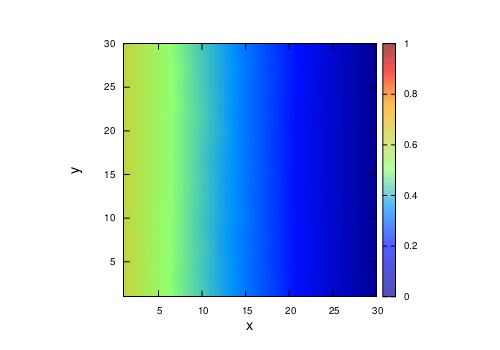}&
\includegraphics[width = 0.25\textwidth]{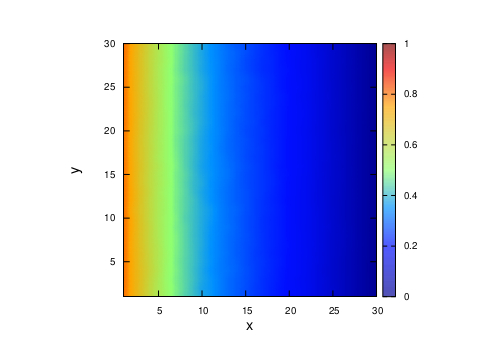}&
\includegraphics[width = 0.25\textwidth]{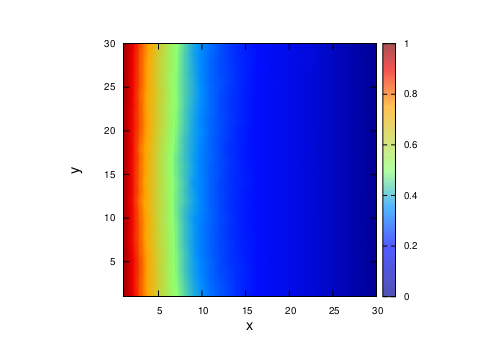}&
\includegraphics[width = 0.25\textwidth]{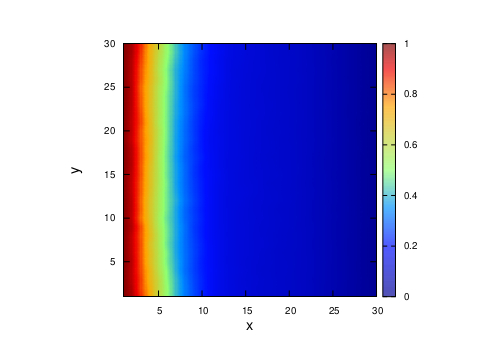}  
\\[0.1cm]
\includegraphics[width = 0.25\textwidth]{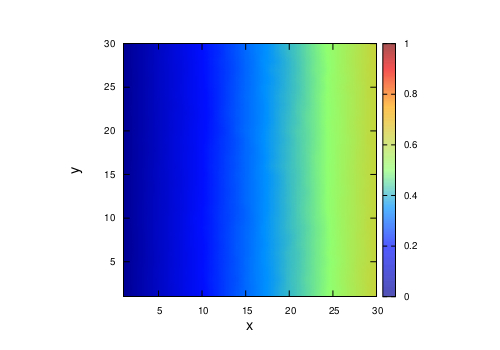}&
\includegraphics[width = 0.25\textwidth]{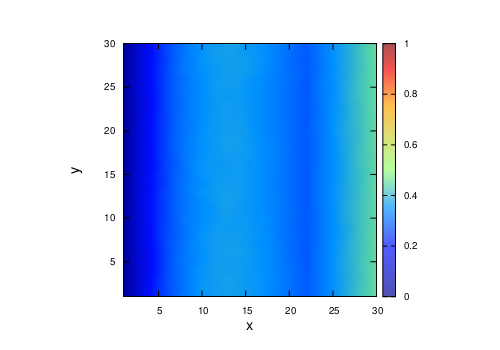}&
\includegraphics[width = 0.25\textwidth]{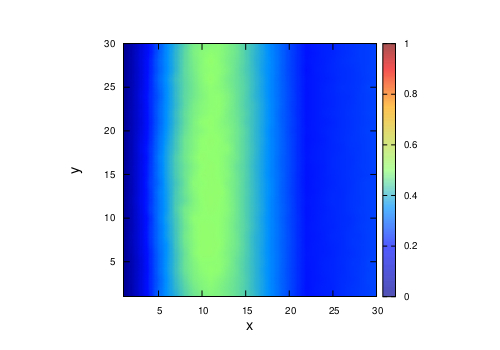}&
\includegraphics[width = 0.25\textwidth]{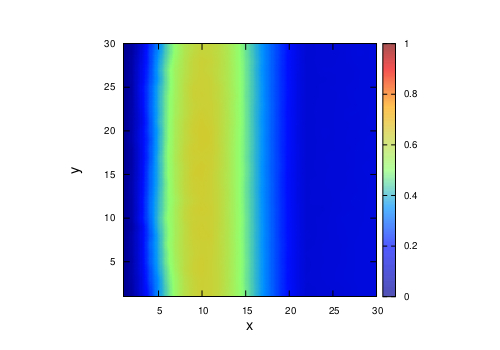} 	
\end{tabular}
\caption{Occupation number profile of passive (top row) and 
active (bottom row) particles at stationarity for  
$L_{\textrm{v}}=23$ and $\varepsilon=0,0.18,0.25,0.35$ (from left to right).
}
\label{fig:fig13}
\end{figure}

The scenario changes drastically for larger values of the 
visibility length, see the bottom panels in 
Figure~\ref{fig:fig3} and, in particular, focus the attention 
on the left one corresponding to $L_{\textrm{v}} = 23$.
Currents are not anymore monotonic functions of the drift, and, 
at very low values of $\varepsilon$, the active 
particle current is higher than the passive particle one, 
as soon as the drift exceeds a certain value the latter overtakes 
the former. Consequently, we see again that by increasing the 
drift of active particles, the transport of passive ones 
is favored. 
This behavior can be explained as before referring to
the occupation number profiles reported in Figure~\ref{fig:fig13}. 
The first three columns are similar to those shown in 
Figure~\ref{fig:fig10} so that the phenomenon 
can be explained similarly. However, in the fourth column
corresponding to $\varepsilon=0.35$, a 
supplementary increase in the occupation number profile of the 
active particles in the middle region of the room is observed
and this explains why for $\varepsilon$ large also the 
passive particle currents becomes negligible. 
In this regime, 
a clogged configuration is eventually reached.

\subsection{Effect of doors}
\label{s:mod2}
\par\noindent
In this section, we discuss the effect of the doors on the dynamics 
of our model. We consider the case in which the doors have equal 
widths 
$w_\textup{L} = w_\textup{R} =14$,
hence their capacity is reduced compared to the corridor model 
described in Subsection~\ref{s:mod1}.
Longitudinal and transversal (i.e., vertical) components 
of the drift, namely, $\varepsilon_1$ and $\varepsilon_2$
are chosen equal and will be simply denoted by $\varepsilon$.

\begin{figure}[t]
\centering
\begin{tabular}{ll}
\includegraphics[width=0.48\textwidth]{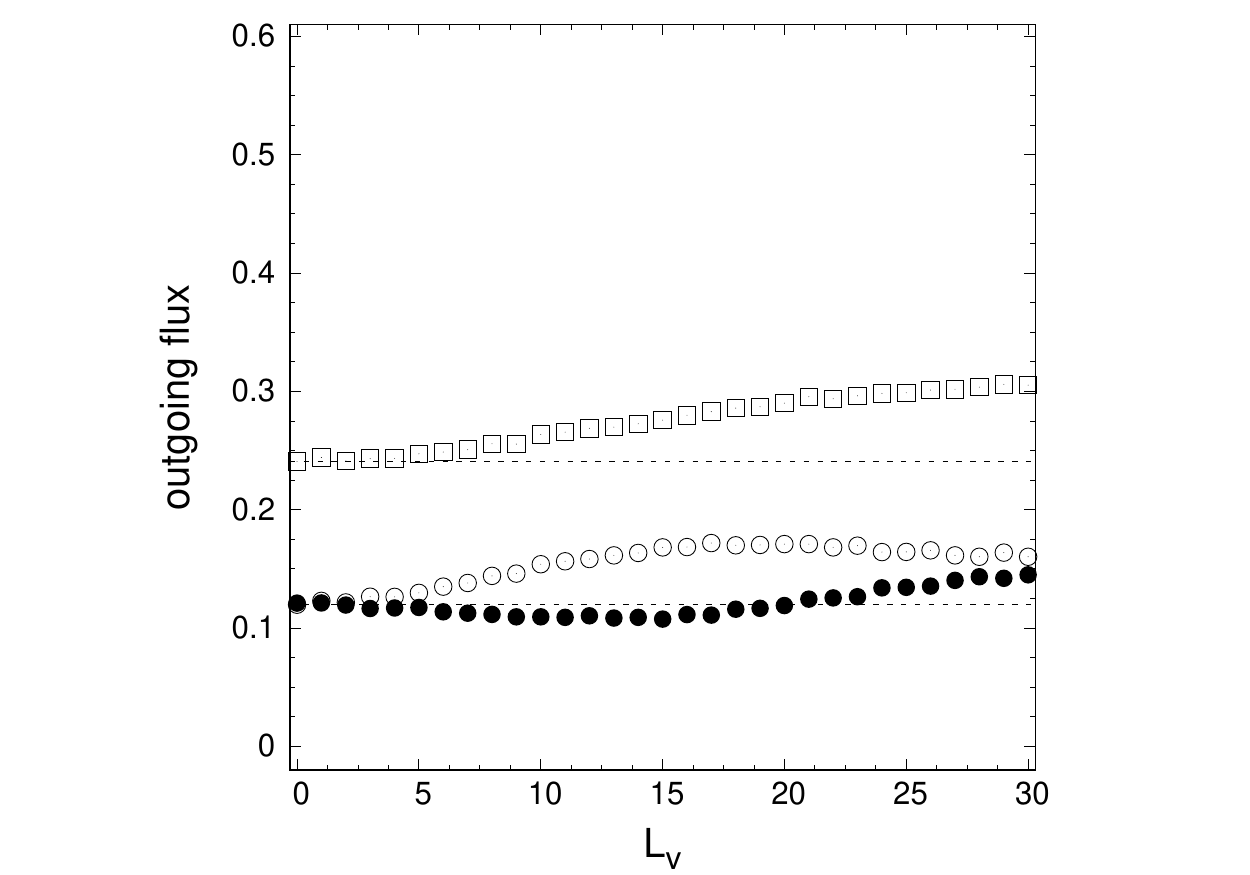}&
\includegraphics[width=0.48\textwidth]{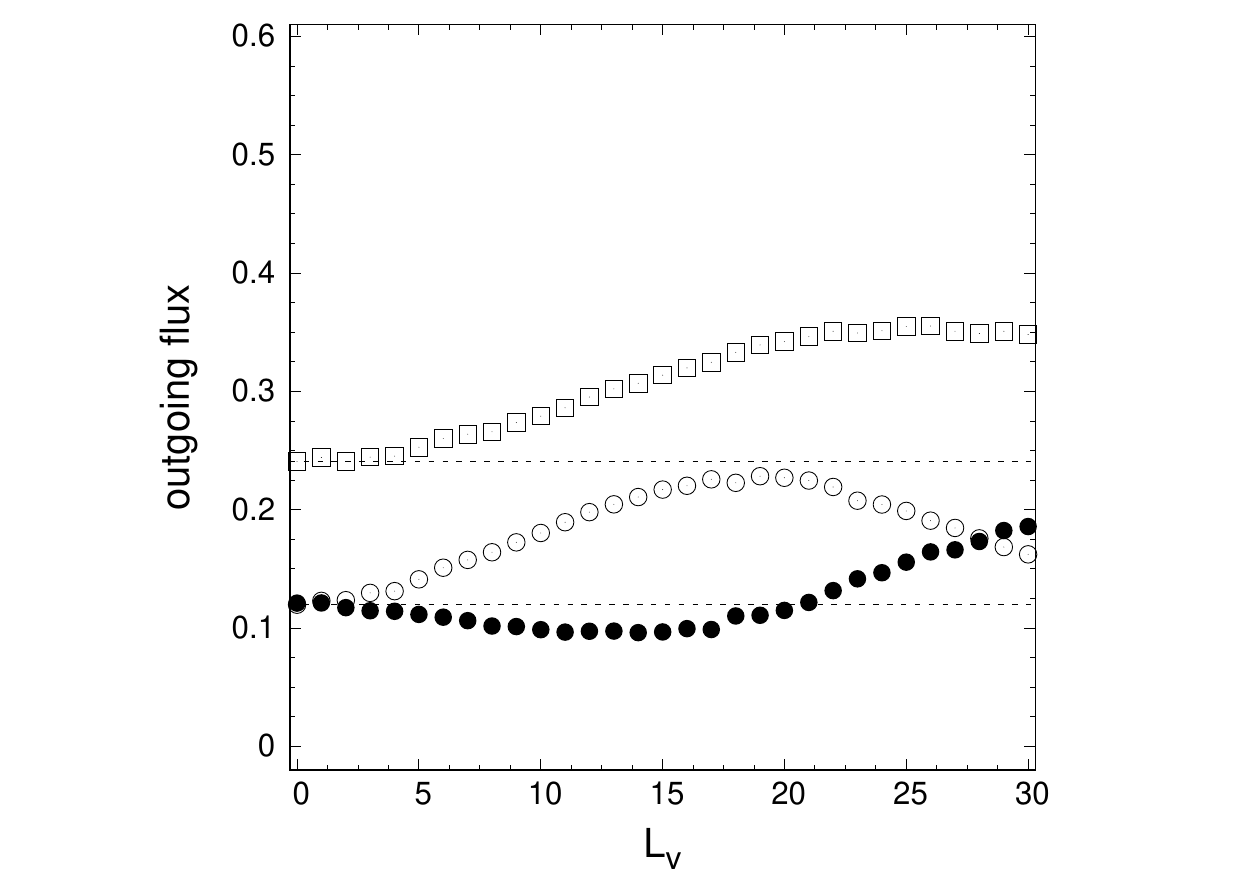}\\[0.1cm]
\includegraphics[width=0.48\textwidth]{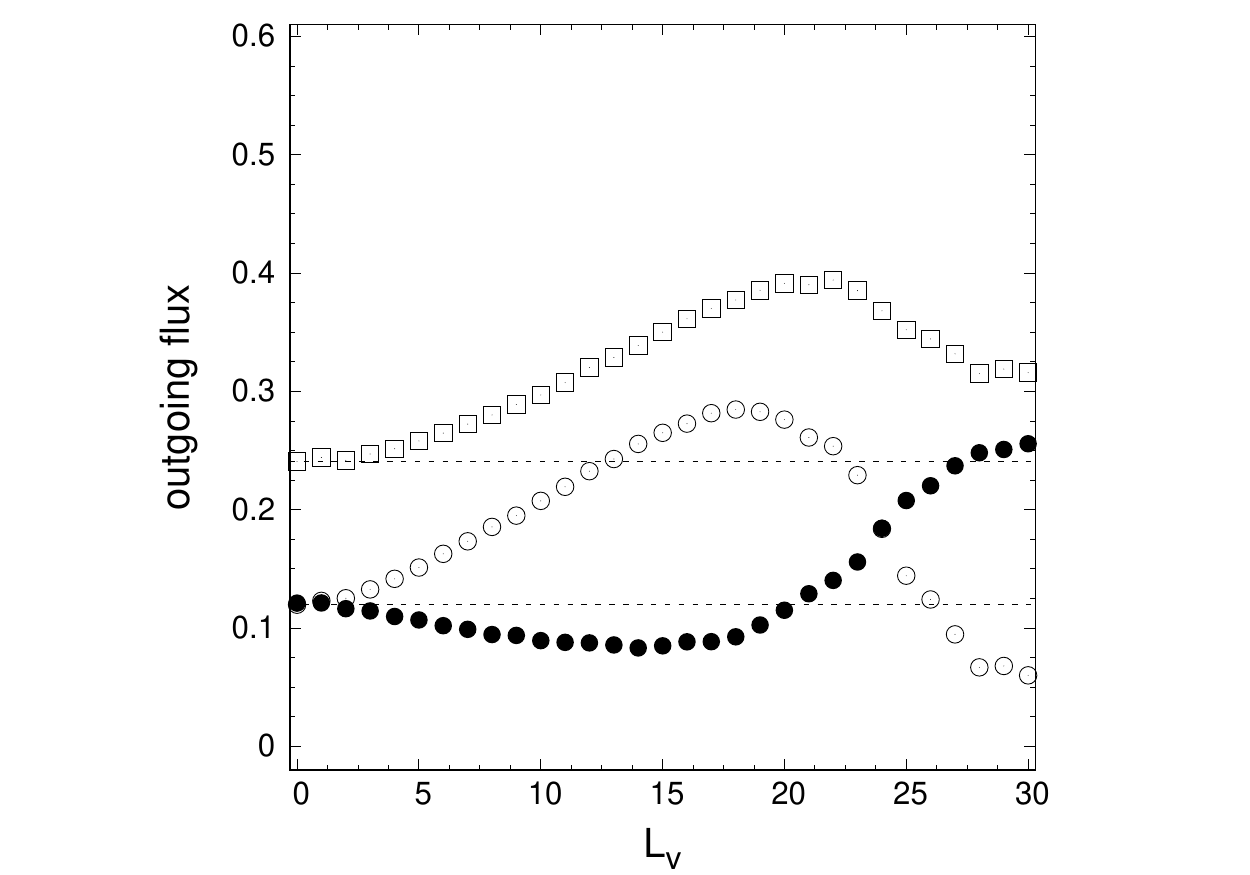}&
\includegraphics[width=0.48\textwidth]{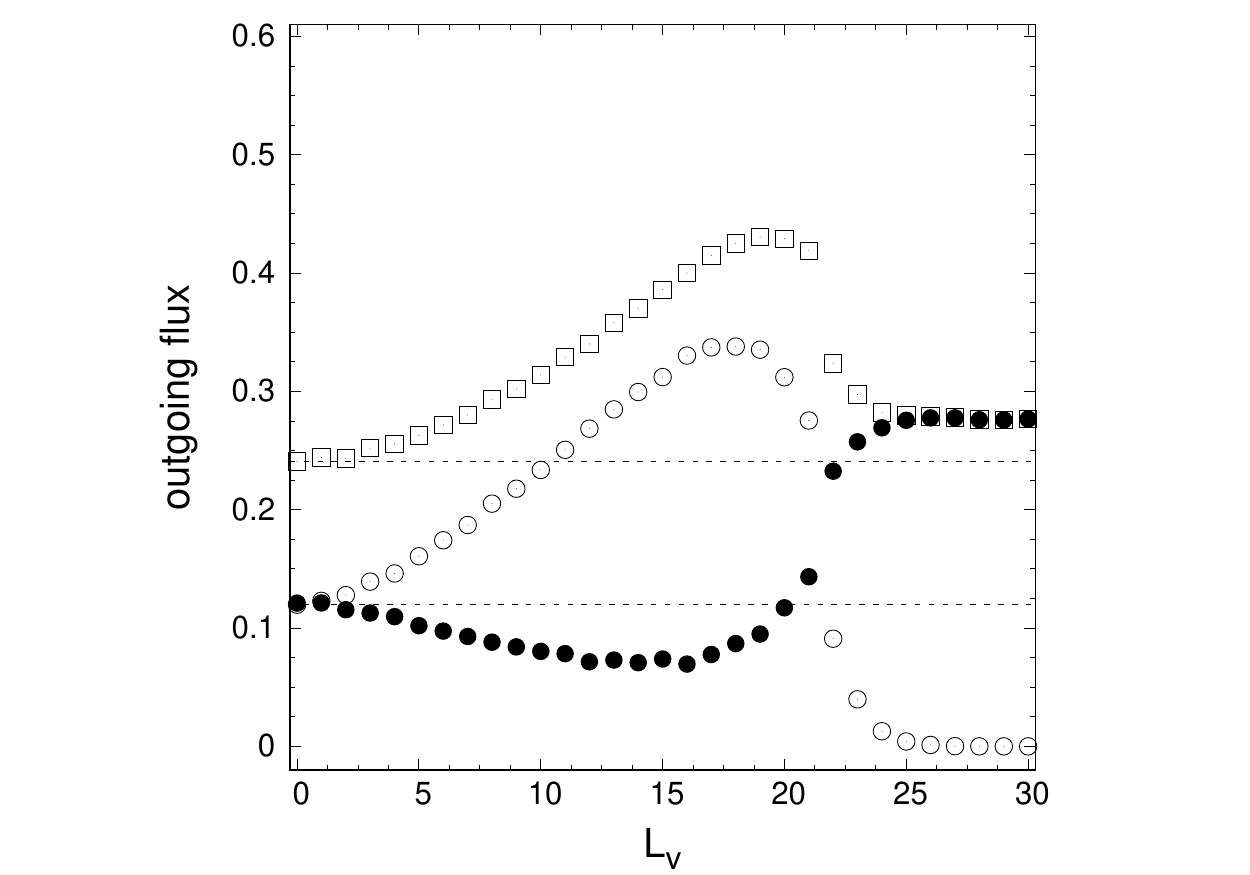}	
\end{tabular}
\caption{Stationary currents of active (empty circles) and passive 
particles (solid disks) and cumulative current (empty squares) as 
functions of $L_{\textrm{v}}$ 
for $\varepsilon=0.05,0.1,0.15,0.2$ (lexicographical order). 
The black dashed lines are as in Figure~\ref{fig:fig1}
}
\label{fig:fig33}
\end{figure}

\begin{figure}[h!]
\centering
\begin{tabular}{lll}
\includegraphics[width = 0.25\textwidth]{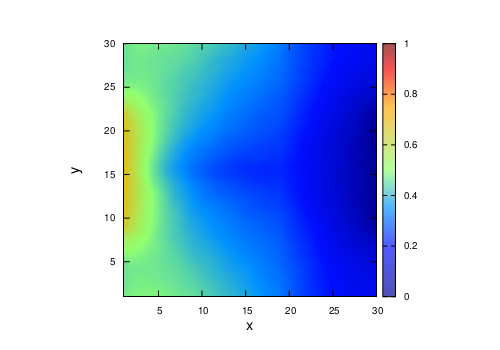}&
\includegraphics[width = 0.25\textwidth]{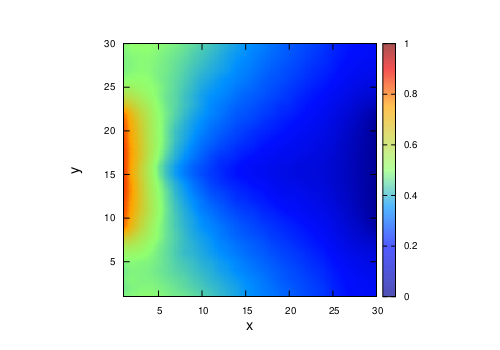}&
\includegraphics[width = 0.25\textwidth]{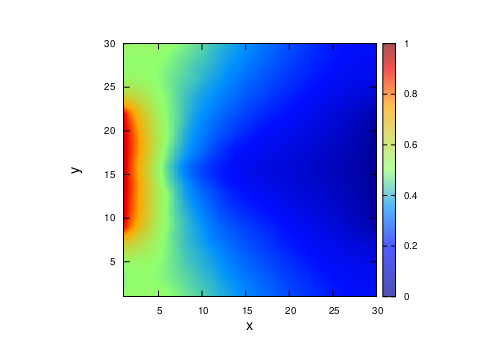}  
\\[0.1cm]
\includegraphics[width = 0.25\textwidth]{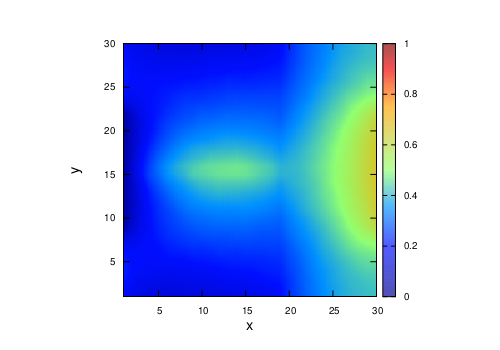}&
\includegraphics[width = 0.25\textwidth]{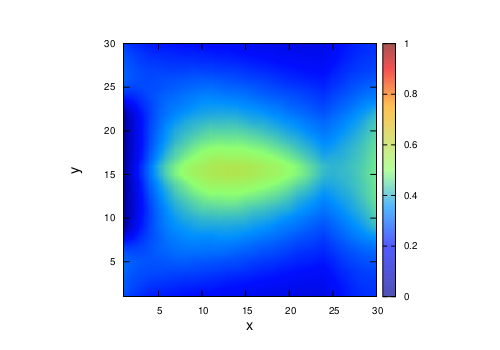}&
\includegraphics[width = 0.25\textwidth]{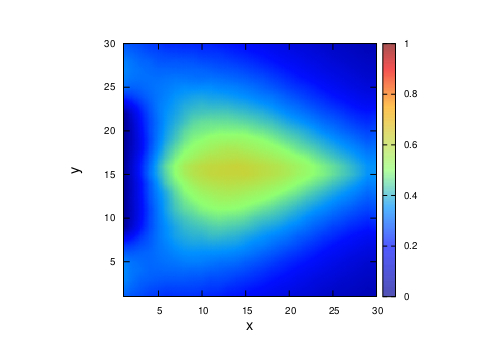}	
\end{tabular}
\caption{Occupation number profile of passive (top row) and 
active (bottom row) particles at stationarity for 
$\varepsilon=0.15$ and $L_{\textrm{v}}=21,25,30$ (from left to right).
}
\label{fig:fig42}
\end{figure}


Since the results are similar to those that we have found in the corridor 
case of Subsection~\ref{s:mod1}, we do not repeat 
the discussion in detail. Instead, we bound ourselves to 
highlight the few key differences that can be observed. 

Figures~\ref{fig:fig33} and \ref{fig:fig42} are analogous 
to Figures~\ref{fig:fig1} and \ref{fig:fig10}. 
The only 
difference is the shape of the region where particles accumulate
which is strongly influenced by the presence of the door and by the 
presence of the transversal drift. The door gives the rounded 
shape to the occupation number profile of active 
particles close to their entrance, whereas the transversal 
drift induces the formation of a ``droplet" in the central region 
of the room. 

\begin{figure}[t]
\centering
\begin{tabular}{ll}
\includegraphics[width=0.48\textwidth]{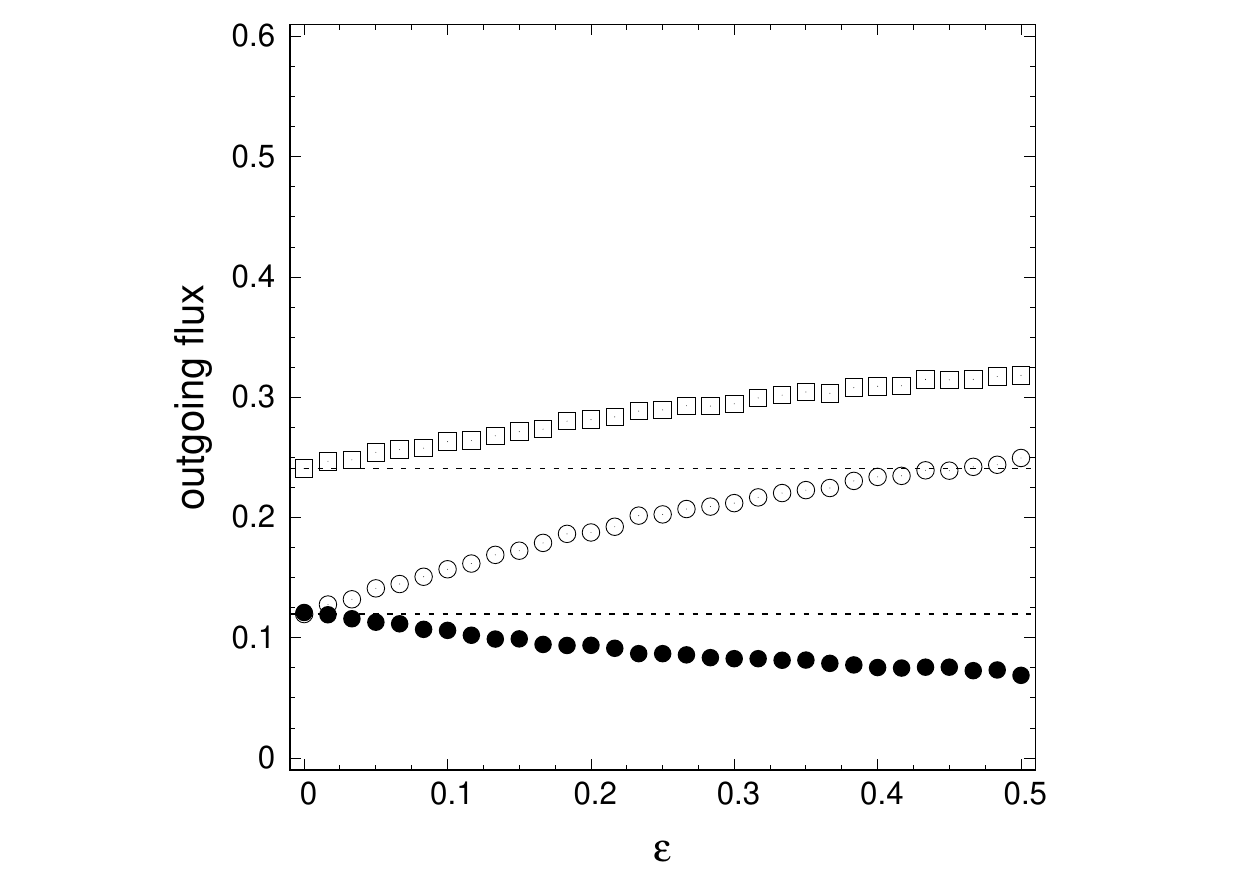}&
\includegraphics[width=0.48\textwidth]{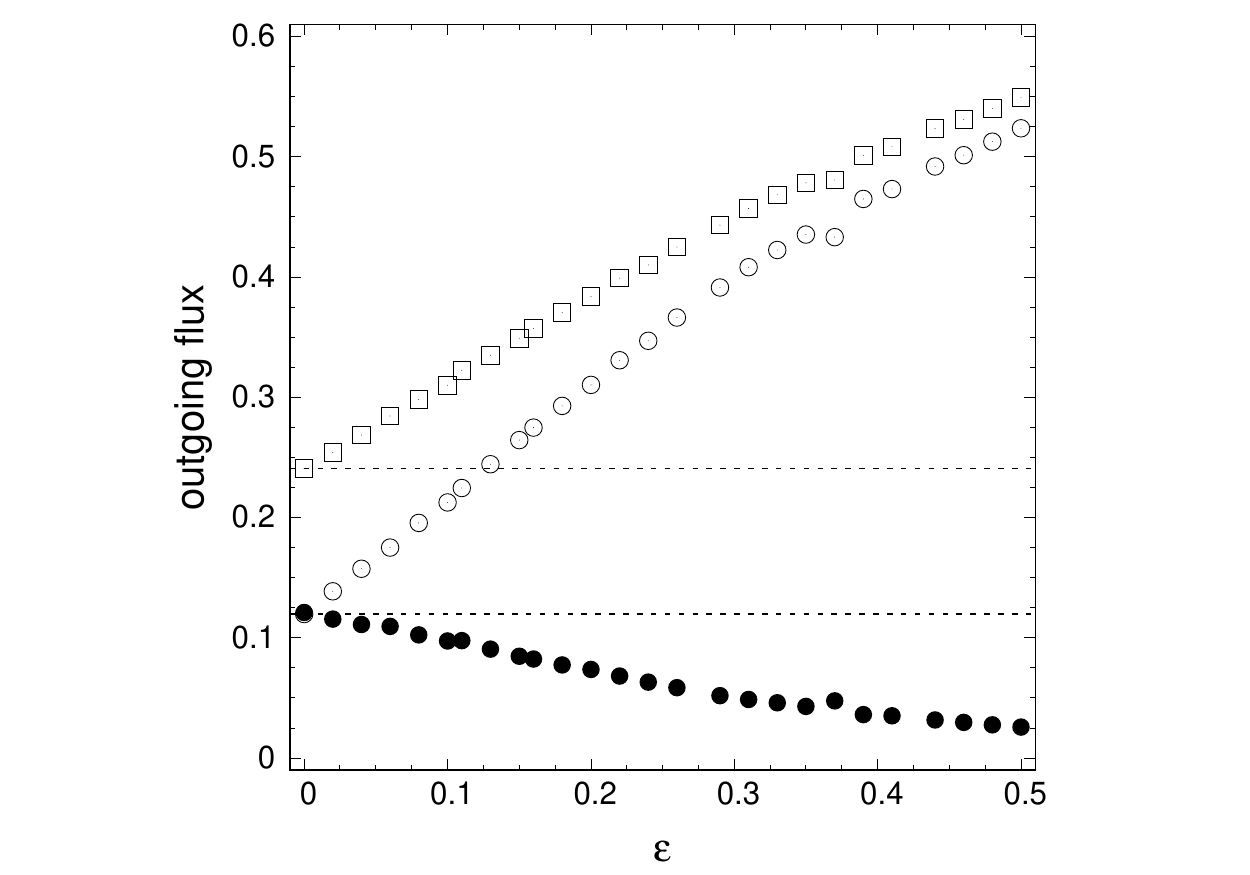}\\[0.1cm]
\includegraphics[width=0.48\textwidth]{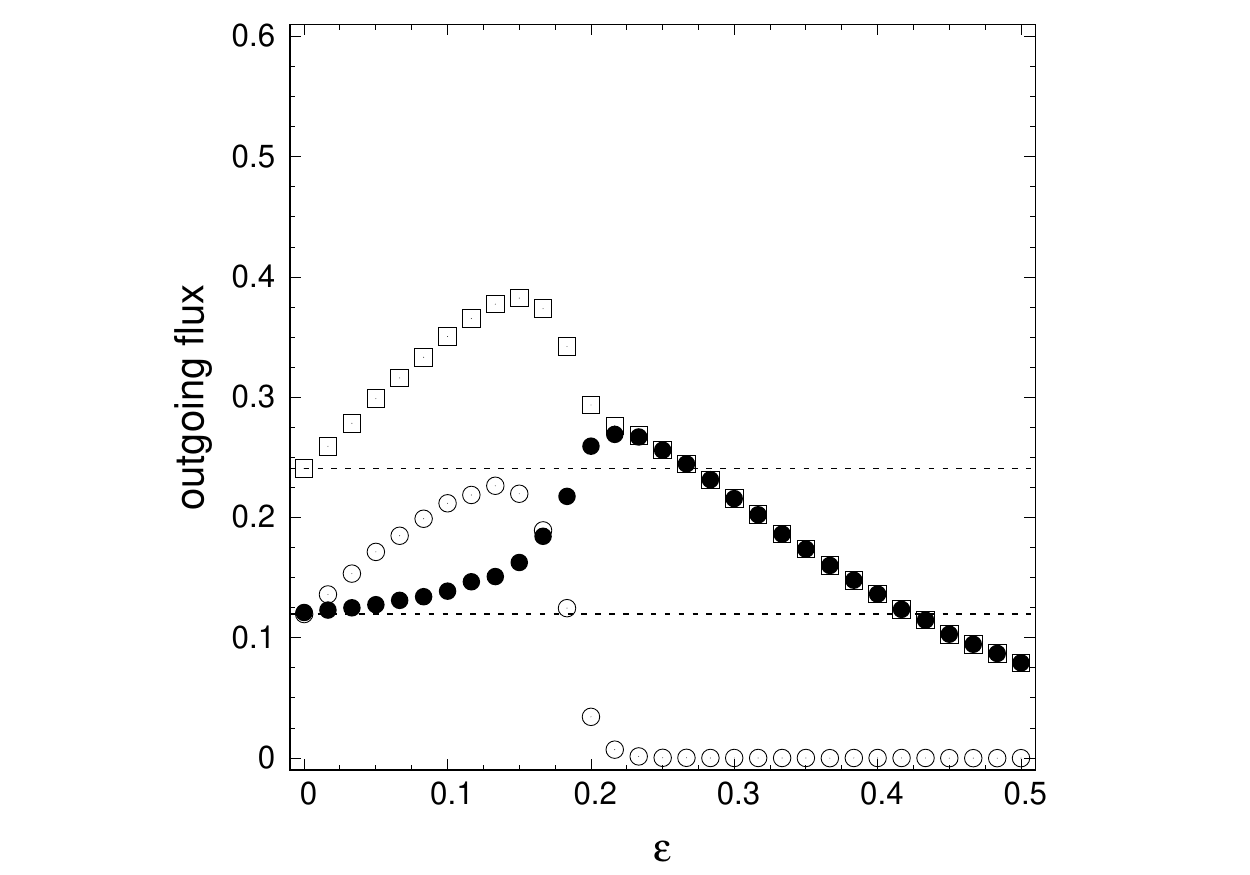}&
\includegraphics[width=0.48\textwidth]{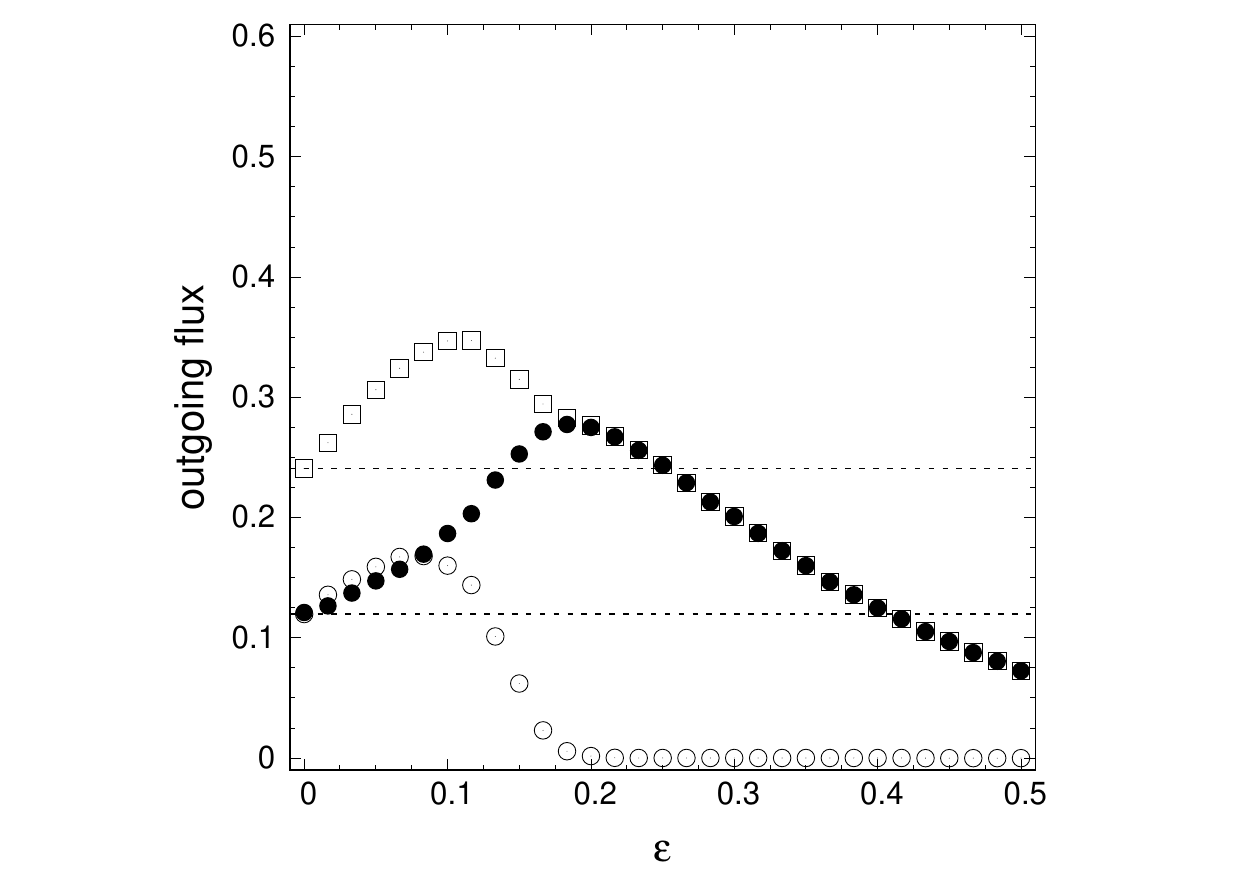}	
\end{tabular}
\caption{Stationary currents of active (empty circles) and passive 
particles (solid disks) and cumulative current (empty squares) as 
functions of $\varepsilon$ for $L_{\textrm{v}}=7,15,23,30$ 
(lexicographical order). 
The black dashed lines are as in Figure~\ref{fig:fig1}
}
\label{fig:fig34}
\end{figure}

\begin{figure}[h!]
\centering
\begin{tabular}{llll}
\includegraphics[width = 0.25\textwidth]{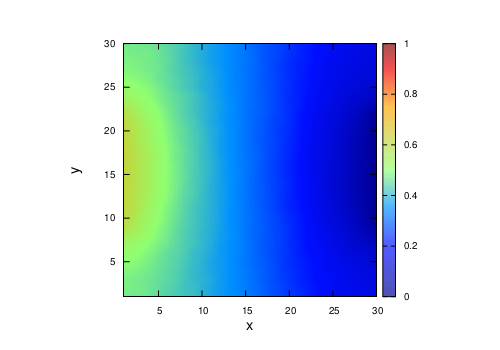}&
\includegraphics[width = 0.25\textwidth]{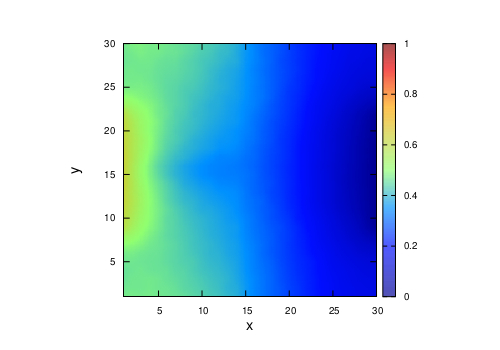}&
\includegraphics[width = 0.25\textwidth]{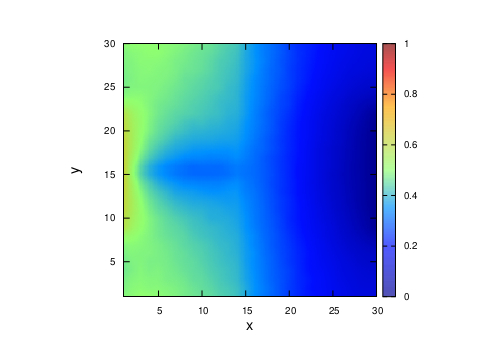}&
\includegraphics[width = 0.25\textwidth]{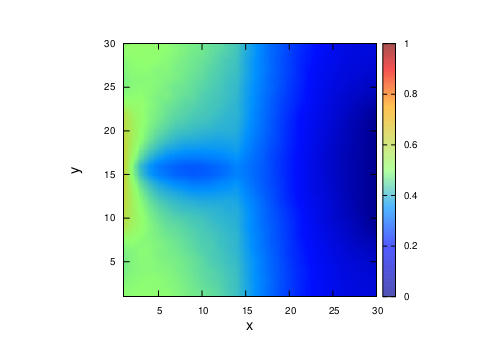}  
\\[0.1cm]
\includegraphics[width = 0.25\textwidth]{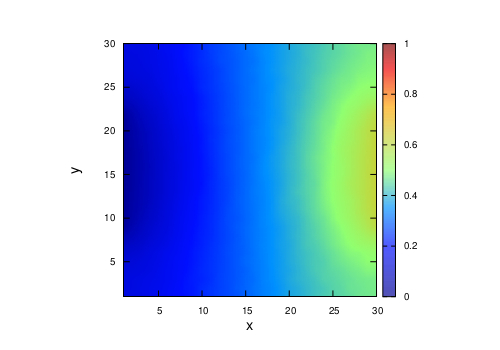}&
\includegraphics[width = 0.25\textwidth]{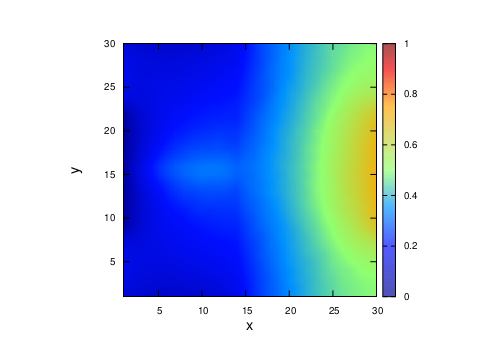}&
\includegraphics[width = 0.25\textwidth]{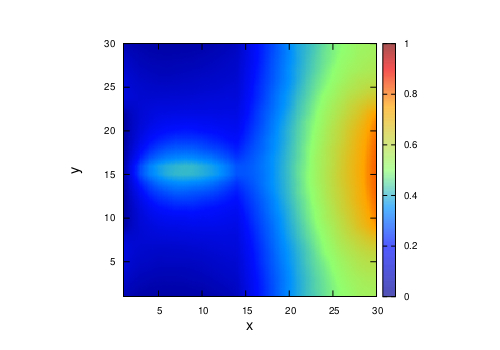}&
\includegraphics[width = 0.25\textwidth]{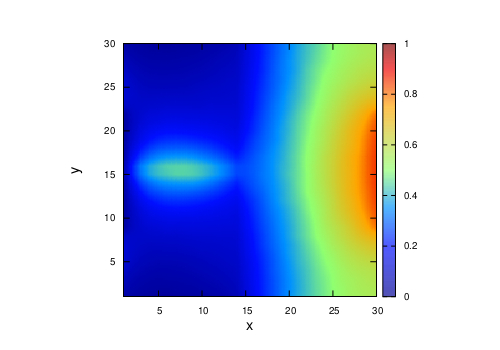} 
\end{tabular}
\caption{Occupation number profile of passive (top row) and 
active (bottom row) particles at stationarity for 
$L_{\textrm{v}}=15$ and 
$\varepsilon=0,0.15,0.35,0.45$ (from left to right).	
}
\label{fig:fig44}
\end{figure}

\begin{figure}[h!]
\centering
\begin{tabular}{llll}
\includegraphics[width = 0.25\textwidth]{hm2ex30new-00-1U.jpg}&
\includegraphics[width = 0.25\textwidth]{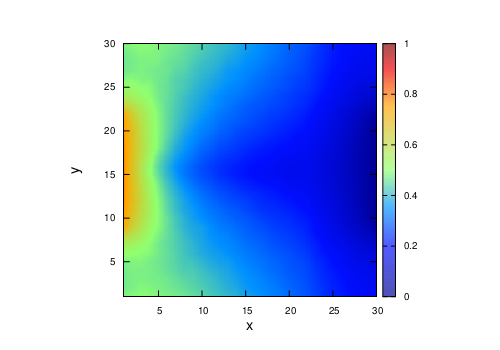}&
\includegraphics[width = 0.25\textwidth]{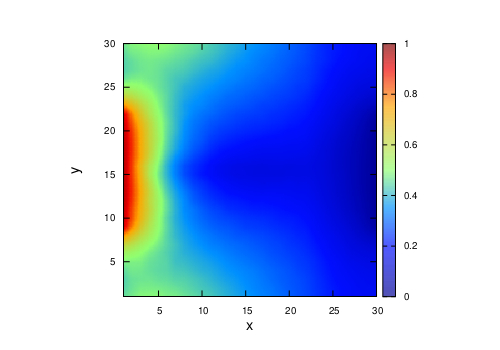}&
\includegraphics[width = 0.25\textwidth]{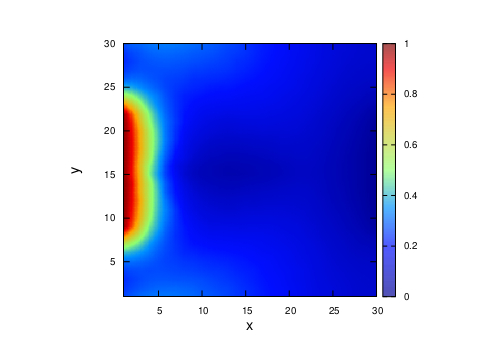}  
\\[0.1cm]
\includegraphics[width = 0.25\textwidth]{hm2ex30new-00-1A.jpg}&
\includegraphics[width = 0.25\textwidth]{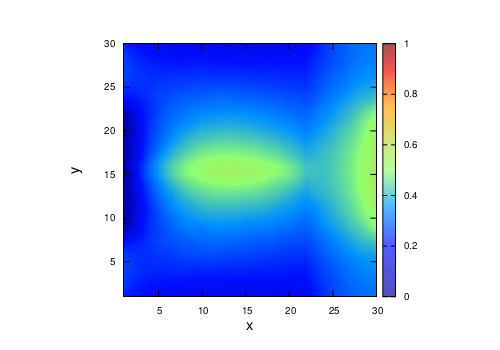}&
\includegraphics[width = 0.25\textwidth]{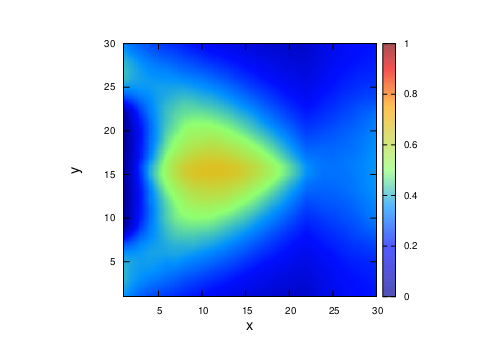}&
\includegraphics[width = 0.25\textwidth]{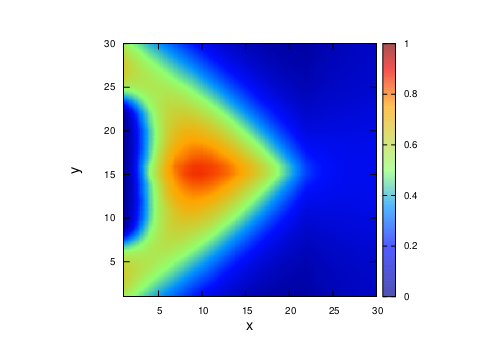} 	
\end{tabular}
\caption{Occupation number profile of passive (top row) and 
active (bottom row) particles at stationarity for 
$L_{\textrm{v}}=23$ and $\varepsilon=0,0.15,0.2,0.3$ (from left to right).
}
\label{fig:fig45}
\end{figure}

Figures~\ref{fig:fig34}--\ref{fig:fig45} are analogous 
to Figures~\ref{fig:fig3}--\ref{fig:fig13}. 
It appears that active particles separate in two distinct
groups. It looks to be a self--induced phase separation within the own 
population. The shape of the droplet is very much affected 
by the geometry of the room and the size of the door. 

\begin{figure}[h!]
\centering
\begin{tabular}{llll}
\includegraphics[width = 0.25\textwidth]{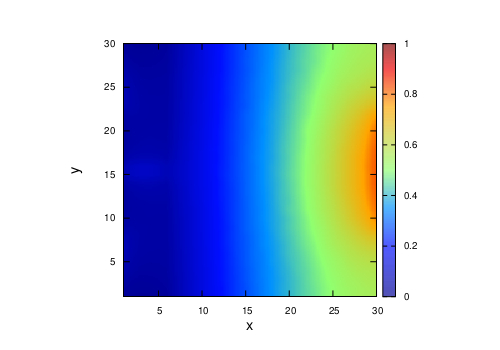}&
\includegraphics[width = 0.25\textwidth]{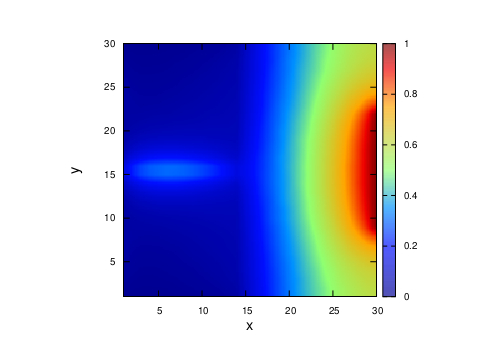}&
\includegraphics[width = 0.25\textwidth]{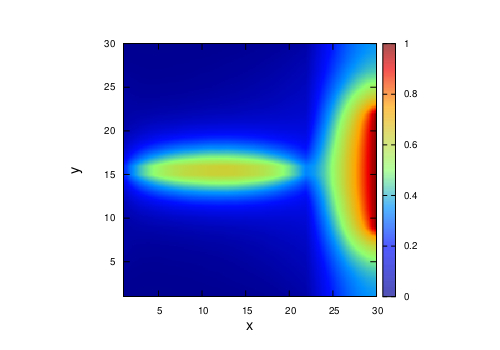}&
\includegraphics[width = 0.25\textwidth]{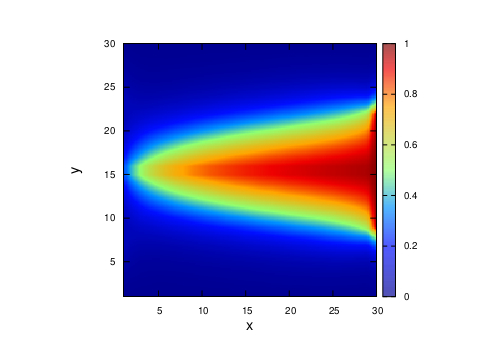} 	
\\[0.1cm]
\includegraphics[width = 0.25\textwidth]{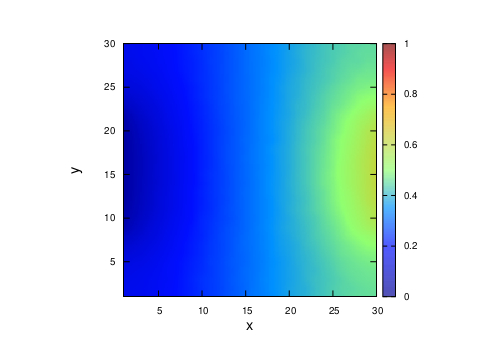}&
\includegraphics[width = 0.25\textwidth]{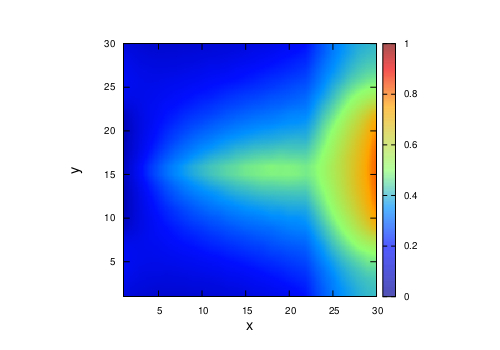}&
\includegraphics[width = 0.25\textwidth]{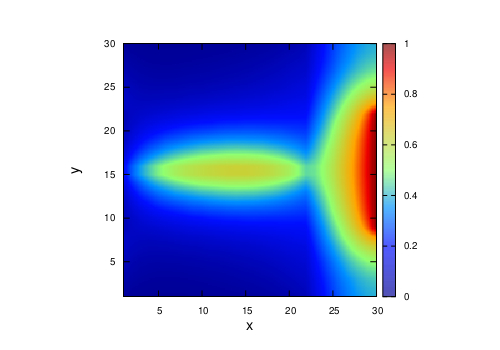}&
\includegraphics[width = 0.25\textwidth]{hm2ex30NEWNP-08-23A.jpg} 	
\end{tabular}
\caption{Occupation number profile at stationarity in a simulation without 
passive particles.
Top row:
$\varepsilon=0.8$ and $L_{\textrm{v}}=7,15,23,30$ (from left to right).
Bottom row: 
$L_{\textrm{v}}=23$ and $\varepsilon=0,0.15,0.5,0.8$ (from left to right).
}
\label{fig:fig46}
\end{figure}

As a final comment, we report that, as we already mentioned 
at the very beginning of Section~\ref{s:res},
we have performed some simulations of the model in absence 
of passive particles, namely, when just active particles 
are present in the lattice. The behavior of the current 
that we find is absolutely standard: the current increases 
monotonically both with respect to the drift and to the  
visibility length.
Nevertheless, we find intersting to show some occupation number 
profiles in Figure~\ref{fig:fig46}, indeed to transverse component 
of the drift, even in this case particles tend ot accumulate in 
the central part of the room. Moreover, depending on the visibility 
length, they can form a central droplet detached 
from the inlet right door. 

\section{Conclusion}
\label{s:con}
\par\noindent
Based on the results detailed in Section~\ref{s:res},
we see that if active particles undergo a non--zero drift 
and the visibility zone is sufficiently large, 
then the outgoing flux of passive particles improves. 
This is essentially the answer to 
the question Q.\ posed in the introduction.
It is due to the fact that, in this regime, active particles
move quickly far from their entrace door. 
In this way their entrance door becomes a free exit for 
the passive particles. 
The dynamics is still slow mainly because active particles 
succeed to jam around the center part of the room, 
slowing down the overall dynamics.

The population of active particles segregates in two 
different structures: one is an agglomeration located 
in the proximity 
of the entrance door, the other is a droplet in the center 
of the visibility zone. This is a consequence of the combined 
action of longitudinal and transversal drifts. On the other hand, 
if the transversal drift is not active, 
we still have an agglomeration in the central 
part of the visibility region,
as we have observed in the 
corridor model.
Its shape is not anymore a droplet but a vertical strip. 

The fact that the flux of passive particles can be controlled 
via the active particle dynamics has been observed for a specific
geometry and for a specific dynamics. The same kind 
of analysis can be done for concrete urban geometries, 
multiple populations of pedestrians, and different dynamics,
providing potentially useful information for large 
crowd management.


\bibliography{mybibfile}

\end{document}